\documentclass[useAMS,usenatbib,A4paper]{mn2e}

\usepackage{times}
\usepackage{natbib}
\usepackage{lscape}
\usepackage[usenames]{color}
\usepackage{graphicx}
\usepackage{amssymb}
\usepackage{wasysym}
\usepackage{pifont}
\usepackage{ulem}

\long\def\Ignore#1{\relax}

\newcommand{\degrees} {^\circ}
\newcommand{\kms}{\mbox{${\rm km\, s^{-1}}$}}
\newcommand{\Msun}{\mbox{$\rm M_{\odot}$}}
\newcommand{\sige}{\mbox{$\sigma_{e}$}}
\newcommand{\sigeight}{\mbox{$\sigma_{e/8}$}}

\newcommand{\Mbh}{\mbox{$\rm M_{\bullet}$}}
\newcommand{\Mbul}{\mbox{$\rm M_{bul}$}}

\newcommand{\Msige}{\mbox{$\rm M_{\bullet}-\sige$}}

\newcommand{\eps}{\mbox{$\epsilon_0$}}
\newcommand{\re}{\mbox{$R_{\rm eff}$}}
\newcommand{\Abar}{\mbox{${\rm A_{bar}}$}}

\newcommand{\D}{\mbox{$\rm \left(D\right)$}}
\newcommand{\B}{\mbox{$\rm \left(B\right)$}}
\newcommand{\BD}{\mbox{$\rm \left(B+D\right)$}}

\newcommand{\eg}{{\it e.g.}}

\newcommand{\ie}{{\it i.e.}}

\newcommand{\cut}[1]{}

\title[The effect of bars on the $\Mbh-\sige$ relation]{The effect of bars on the $\Mbh-\sige$ relation: offset, scatter and residuals correlations}
\author[M. Hartmann et al.]{
Markus Hartmann$^{1,2}$\thanks{E-mail: {\tt hartmann@ari.uni-heidelberg.de}}, 
Victor P. Debattista$^{2}$\thanks{E-mail: {\tt vpdebattista@gmail.com}},
David R. Cole$^{2}$\thanks{E-mail: {\tt drdrcole@gmail.com}},
Monica Valluri$^{3}$\thanks{E-mail: {\tt mvalluri@umich.edu}},
\newauthor
Lawrence M. Widrow$^{4}$\thanks{E-mail: {\tt widrow@astro.queensu.ca}},
Juntai Shen$^{5}$\thanks{E-mail: {\tt jshen@shao.ac.cn}}
\\
$^{1}$Astronomisches Rechen-Institut, Zentrum f\"ur Astronomie der Universit\"at Heidelberg (ZAH), M\"onchhofstr. 12-14, 69120 Heidelberg, Germany\\
$^{2}$Jeremiah Horrocks Institute, University of Central Lancashire, Preston, PR1 2HE, United Kingdom \\
$^{3}$Department of Astronomy, University of Michigan, 500 Church St., Ann Arbor, MI 48109, USA\\
$^{4}$Department of Physics, Engineering PHysics, and Astronomy, Queen's University, 99 University Avenue, Kingston, Ontario K7L 3N6, Canada\\
$^{5}$Shanghai Astronomical Observatory, 80 Nandan Road, Shanghai 200030, China}
\begin{document}

\date{Accepted xxx Received xxx ; in original form \today}

\maketitle

\label{firstpage}


\begin{abstract}
  We analyse a set of collisionless disc galaxy simulations to study
  the consequences of bar formation and evolution on the \Msige\
  relation of supermassive black holes. The redistribution of angular
  momentum driven by bars leads to a mass increase within the central
  region, raising the velocity dispersion of the bulge, \sige, on
  average by $\sim 12\%$ and as much as $\sim20\%$.  If a disc galaxy
  with a SMBH satisfying the \Msige\ relation forms a bar, and the
  SMBH does not grow in the process, then the increase in \sige\ moves
  the galaxy off the \Msige\ relation.  We explore various effects
  that can affect this result including contamination from the disc
  and anisotropy.  The displacement from the \Msige\ relation for
  individual model barred galaxies correlates with both ${\rm M\B /
    M\BD}$ and $\beta_\phi\BD$ measured within the effective radius of
  the bulge.  Overall, this process leads to an \Msige\ for barred
  galaxies offset from that of unbarred galaxies, as well as an
  increase in its scatter.  We assemble samples of unbarred and barred
  galaxies with classical bulges and find tentative hints of an offset
  between the two consistent with the predicted.  Including all barred
  galaxies, rather than just those with a classical bulge, leads to a
  significantly larger offset.
\end{abstract}


\section{Introduction}  
\label{sec:intro} 

One of the most striking results to emerge from {\it Hubble Space
  Telescope} observations of galactic nuclei is that essentially every
galaxy with a significant stellar spheroid contains a supermassive
black hole (SMBH) whose mass is correlated with properties of the host
galaxy. For instance the masses of SMBHs, \Mbh, are found to correlate
with the bulge luminosity, $\rm{L_{bul}}$ \citep{Kormendy1995,
  Marconi2003, Graham2007a, Gueltekin2009b, Sani2011, McConnell2011,
  Beifiori2012, Graham2013}, with the bulge mass, $\Mbul$
\citep{Magorrian1998, Marconi2003, Haering2004, Sani2011,
  Beifiori2012}, the bulge velocity dispersion, $\sige$
\citep{Gebhardt2000, Ferrarese2000, Merritt2001, Tremaine2002,
  Ferrarese2005, Gueltekin2009b, McConnell2011, Graham2011,
  Beifiori2012}, with the mass of the galaxy, $\rm M_{gal}$
\citep{Ferrarese2002, Baes2003, Kormendy2011b, Volonteri2011,
  Beifiori2012}, with the S\'{e}rsic index of the surface brightness
profile, $n$ \citep{Graham2007}, with the spiral pitch angle
\citep{Seigar2008, Berrier2013}, with the number of globular clusters
\citep{Burkert2010, Harris2011, Rhode2012}, with the globular cluster
system velocity dispersion \citep{Sadoun2012, Pota2013}, and with the
inner core radius, $r_\gamma$ \citep{Lauer2007a, Kormendy2009}.
Amongst these, the \Msige\ relation with the form
$\log{\left(\Mbh/\Msun\right)} = \alpha+\beta
\log{\left(\sigma/200\kms\right)}$ \citep{Gebhardt2000, Ferrarese2000}
is one of the tightest \citep{Gebhardt2003, Marconi2003,
  Gueltekin2009b} though the scatter has increased in recent studies
(see Table~\ref{tab:slopes}).  Here $\beta$ is the slope and $\alpha$
is the zero-point of the relation.
Measurements of $\beta$ have produced a variety of different results
(see Table~\ref{tab:slopes}).  Early estimates varied from
$3.75\pm0.3$ \citep{Gebhardt2000} to $4.8\pm0.5$
\citep{Ferrarese2000}.  More recently \citet{Gueltekin2009b} found
$\beta = 4.24\pm0.41$, whereas \citet{McConnell2013} found $\beta =
5.64\pm0.32$ and \citet{Graham2011} found $\beta = 5.13\pm0.34$,
demonstrating that the slope of the relation remains imperfectly
defined.  Two sources of this variation are the uncertainty in the
data (see for example Section 5 of \cite{McConnell2013}) and different
slopes in different galaxy types.  \cite{McConnell2013} find different
values of $\alpha$ and $\beta$ for early and late type galaxies (see
Table~\ref{tab:slopes}) while \citet{Graham2011} and
\citet{Graham2013} find different values of $\alpha$ and $\beta$ for
barred and unbarred galaxies. Such differences must be explained by
any model explaining the link between SMBHs and their hosts.

\begin{table}
  \begin{center}
    \begin{tabular}{lccc} 
      \hline
      \multicolumn{1}{l}{} &
      \multicolumn{1}{c}{$\alpha$} &
      \multicolumn{1}{c}{$\beta$} &
      \multicolumn{1}{c}{scatter} \\[-0.5ex]
      \hline
\citet{Gebhardt2000} & $8.08\pm0.06$ & $3.75\pm0.3$ & 0.3 \\
\citet{Merritt2001} & $8.11\pm0.11$ & $4.72\pm0.36$ & 0.35 \\
\citet{Tremaine2002} & $8.13\pm0.06$ & $4.02\pm0.32$ & 0.33 \\
\citet{Ferrarese2005} & $8.22\pm0.06$ & $4.86\pm0.43$ & 0.34 \\
\citet{Gueltekin2009b} & $8.12\pm0.08$ & $4.24\pm0.41$ & 0.44 \\
\citet{Graham2011} & $8.13\pm0.05$ & $5.13\pm0.34$ & 0.43 \\
\citet{Beifiori2012} & $8.19\pm0.07$ & $4.17\pm0.32$ & 0.41 \\
\citet{McConnell2013} & $8.32\pm0.05$ & $5.64\pm0.32$ & 0.38 \\
\hline
\hline
\citet{Graham2011} \\
barred & $7.80\pm0.10$ & $4.34\pm0.56$ & 0.36 \\
unbarred & $8.25\pm0.06$ & $4.57\pm0.35$ & 0.37 \\
elliptical & $8.27\pm0.06$ & $4.43\pm0.57$ & 0.34 \\
      \hline
\citet{McConnell2013} \\
early type & $8.07\pm0.21$ & $5.20\pm0.36$ &  \\
late type & $8.39\pm0.06$ & $5.06\pm1.16$ &  \\
      \hline
    \end{tabular}
  \end{center}
  \caption[]{\label{tab:slopes} Published values for the zero-point, $\alpha$, slope, $\beta$,
    and scatter of the \Msige\ relation from a number of studies. For the \citet{Graham2011} and \citet{McConnell2013} studies, we also present the results found for different galaxy type. }
\end{table}

These scaling relations suggest that there is a connection between the
growth of the SMBH and the bulge.  However the causal basis of these
scaling relations is still not fully understood.  Does the presence of
a SMBH govern the bulge's growth or is the growth of the SMBH
determined by the bulge it resides in? The vast energy available from
an accreting SMBH during its phase as an active galactic nucleus (AGN)
can couple the SMBH to its host, since only a small fraction of this
energy is needed to alter the temperature and structure of the
surrounding interstellar medium \citep{Silk1998, King2003, Wyithe2003,
  DiMatteo2005, Murray2005, Sazonov2005, Younger2008, Booth2009,
  Power2011}.  Alternatively, the \Msige\ relation could merely be a
consequence of the merger history in a hierarchical universe
\citep{Adams2001, Adams2003, Volonteri2009, Jahnke2011}.

\citet{Graham2008a} and \citet{Graham2009} found that SMBHs in barred
galaxies have an offset from the \Msige\ relation of elliptical
galaxies \citep[see also][]{Graham2011, Graham2013}. In addition
excluding barred galaxies from the \Msige\ relation reduces the
scatter $\eps$ from 0.47 to 0.41 \citep{Graham2009}.  Both
\citet{Hu2008} and \citet{Gadotti2009} point out that the presence of
bars could be responsible for the difference in these \Msige\
relations.  \citet{Graham2008b} and \citet{Graham2009} obtained a
\Msige\ relation for unbarred galaxies and \citet{Graham2011} showed
that barred galaxies have an offset of $\sim0.5$~dex from this
relation. On the other hand, amongst active galaxies with
$\Mbh<2\times10^6$~\Msun, \citet{Xiao2011} found no significant offset
of barred galaxies relative to the \Msige\ relation of unbarred
galaxies.  Likewise, in a sample of galaxies with active nuclei for
which they obtained upper limits on $\Mbh$, \citet{Beifiori2009} found
no systematic difference between barred and unbarred galaxies.

Bars, either weak or strong, are present in $\sim 65\%$ of local
luminous disc galaxies \citep{Knapen1999, Eskridge2000, Nair2010,
  Masters2011}.  The fraction of {\it strongly} barred galaxies rises
from $\sim 20\%$ at $z\sim 1$ to $\sim 30\%$ at $z=0$
\citep{Elmegreen2004, Jogee2004a, Sheth2008, Skibba2012}. Thus bars
have had a long time to drive evolution in disc galaxies
\citep{Courteau1996, Debattista2004, Kormendy2004, Jogee2005,
  Debattista2006}.  Bars lead to a redistribution of angular momentum
and an increase in the central mass density \citep{Hohl1971}.
Therefore they provide a possible mechanism for fuelling central
starbursts and AGN activity \citep{Simkin1980, Athanassoula1992b,
  Shlosman1989,Jogee2005, Schawinski2011, Hicks2013}.  While near
infrared surveys find no difference in the fraction of barred galaxies
between active and non-active galaxies \citep{McLeod1995,
  Mulchaey1997}, this could be due to the vastly disparate timescales
involved, with AGN having an active phase of order $10^6$~Myr compared
to the $\sim100\times$ longer quiescent phase \citep{Shabala2008}.

Besides fuelling the SMBH, bars may affect a galaxy's position on the
\Msige\ relation in other ways.  \citet{Graham2011} proposed that
several bar driven effects can cause offsets in the \Msige\ relation,
including velocity anisotropy, and the increase in \sige\ due to mass
inflows, angular momentum redistribution and buckling.  Bars can also
lead to the growth of pseudo bulges by driving gas to the centre to
fuel star formation \citep{Kormendy2004}, changing \sige.  Lastly, by
transferring angular momentum outwards \citep{Lynden-Bell1972,
  Tremaine1984, Debattista2000, Athanassoula2002, Athanassoula2003,
  Sellwood2006, Berentzen2007}, bars increase the central density of
the disc, raising the velocity dispersion of the bulge
\citep{Debattista2005, Debattista2013}.

This paper explores the effect of bar evolution on the \Msige\
relation of classical bulges, assuming that they form with a SMBH
satisfying the \Msige\ relation and later the disc develops a bar.  We
show, using collisionless simulations, that as a result of angular
momentum redistribution, such a SMBH ends up offset from the \Msige\
relation.  In a companion paper \citet{Brown2013} examine the effect
of the growth of a SMBH on the nuclear stellar kinematics in both pure
disc systems, and in systems composed of a disc and spheroidal bulge.
They show that the presence of a bar enhances the effect that the
growth of an SMBH has on the stellar \sige.  Their simulations show
that the growth of a SMBH after the formation of a bar also causes an
offset in \sige, but one that is smaller than that resulting from the
formation and evolution of a bar.  Thus the current paper and the
\citet{Brown2013} paper show that regardless of whether the SMBH
exists prior to bar formation or whether it grows after bar formation
(with reality being somewhere in between these two extremes), barred
galaxies will have larger values of \sige\ than unbarred galaxies with
the same \Mbh.

The paper is organised as follows: In Section~\ref{sec:sim} we
describe the simulations.  In Section~\ref{sec:bar} we study what
effect bar formation and evolution have on the bulge and disc and how
this might effect the \Msige\ relation.  We discuss the consequences
of bar evolution for the \Msige\ relation in Section~\ref{sec:msigma}
and compare our results with the observed \Msige\ relations of
classical bulges in unbarred and barred galaxies.
Section~\ref{sec:conclusions} sums up our findings.


\begin{landscape}
\begin{table}
\begin{minipage}{230mm}
  \caption{The sample of disc galaxy simulations used in this study.
    In the left column we list the run number and some of the initial
    parameters of each simulation: the minimum of the Toomre $Q$, the
    minimum of the swing amplification parameter $X$, the
    disc-to-bulge ratio D/B, and the halo-to-bulge ratio H/B, within
    \re\ (obtained by calculating the projected radius containing half
    the mass of the bulge), and the S\'{e}rsic index $n$ of the bulge.
    In the right part of the table we show the parameters of the
    evolved system: the bar amplitude \Abar\ at $t_1$ and $t_2$, \re\
    at $t_0=0$, $t_1$ and $t_2$, the fractional change in mass
    $\Delta{\rm M\BD / M\BD_{init}}$ within \re\ at $t_1$ and $t_2$,
    the aperture velocity dispersion \sige\ of bulge$+$disc particles
    measured within a circular aperture at $t_0$, $t_1$ and $t_2$ and
    the dispersion scatter $\Delta\sige$ of bulge$+$disc particles at
    $t_2$.  Simulations 16 and 21 are very similar in their setup;
    coincidentally, the effects of stochastically \citep{Sellwood2009,
      Roskar2012} are weak in these two baryon-dominated simulations.
  }
\begin{center}
\begin{tabular}{cccccc|@{\vline}|ccccccccccc@{}}
\hline
\hline
Run & Q & X & D/B & H/B & $n$ & \Abar\  & \Abar\  & \re$\left(t_0\right)$ & \re$\left(t_1\right)$ & \re$\left(t_2\right)$ & $\frac{\Delta \rm M \left(B+D\right)}{\rm M \left(B+D\right)_{init}}$ & $\frac{\Delta \rm M \left(B+D\right)}{\rm M \left(B+D\right)_{init}}$ & $\sige\left(t_0\right)$ & $\sige\left(t_1\right)$ & $\sige\left(t_2\right)$ & $\Delta\sige\left(t_2\right)$ \\
 & & & & & & $\left(t_1\right)$ & $\left(t_2\right)$ & [pc] & [pc] & [pc] & $\left(t_1\right)$ & $\left(t_2\right)$ &  [\kms] & [\kms] & [\kms] & [\kms] \\ \hline
%
 1  & 1.02 & 2.58 & 5.8 & 0.02 & 1.0 & 0.140 & 0.134 & 593 & 489 & 494 & 0.31 & 0.34 & 102.0 & 142.0 & 144.4 &  7.1 \\      
 2  & 1.01 & 2.98 & 4.8 & 0.05 & 1.3 & 0.176 & 0.180 & 659 & 570 & 572 & 0.24 & 0.29 & 102.5 & 137.4 & 142.0 &  9.4 \\
 3  & 1.00 & 3.41 & 4.5 & 0.04 & 1.7 & 0.117 & 0.167 & 649 & 599 & 599 & 0.21 & 0.28 & 108.9 & 136.7 & 144.8 &  9.6 \\  
 4  & 1.04 & 3.71 & 3.8 & 0.06 & 1.5 & 0.177 & 0.247 & 751 & 682 & 665 & 0.19 & 0.27 & 103.7 & 129.6 & 140.8 & 10.9 \\  
 5  & 1.13 & 4.41 & 3.6 & 0.11 & 1.7 & 0.116 & 0.219 & 780 & 748 & 727 & 0.12 & 0.21 & 105.9 & 120.1 & 133.0 &  9.5 \\  
 6  & 1.27 & 2.61 & 5.9 & 0.02 & 1.2 & 0.222 & 0.270 & 649 & 561 & 552 & 0.25 & 0.34 & 103.7 & 135.8 & 145.3 & 12.5 \\  
 7  & 1.25 & 2.99 & 5.0 & 0.03 & 1.6 & 0.138 & 0.212 & 610 & 542 & 541 & 0.23 & 0.30 & 105.3 & 133.5 & 140.9 &  9.6 \\ 
 8  & 1.25 & 3.51 & 4.8 & 0.15 & 1.3 & 0.149 & 0.220 & 596 & 537 & 536 & 0.18 & 0.26 & 107.4 & 129.8 & 140.9 & 10.3 \\ 
 9  & 1.24 & 3.95 & 3.6 & 0.08 & 1.7 & 0.111 & 0.178 & 738 & 726 & 704 & 0.10 & 0.21 & 107.2 & 116.2 & 131.1 &  7.8 \\  
 10 & 1.26 & 4.46 & 3.7 & 0.18 & 1.8 & 0.006 & 0.139 & 752 & 774 & 748 & 0.02 & 0.12 & 108.9 & 108.0 & 120.9 &  7.1 \\  
 11 & 1.41 & 2.51 & 6.3 & 0.02 & 1.4 & 0.121 & 0.137 & 456 & 440 & 440 & 0.17 & 0.25 & 115.2 & 132.8 & 142.7 &  6.9 \\  
 12 & 1.50 & 3.03 & 4.8 & 0.03 & 1.3 & 0.164 & 0.134 & 531 & 495 & 498 & 0.14 & 0.18 & 109.4 & 128.2 & 131.9 &  6.2 \\ 
 13 & 1.50 & 3.51 & 6.4 & 0.20 & 1.0 & 0.260 & 0.265 & 727 & 656 & 633 & 0.21 & 0.29 &  94.6 & 117.4 & 126.4 & 11.0 \\  
 14 & 1.50 & 4.00 & 5.5 & 0.26 & 1.0 & 0.176 & 0.267 & 841 & 732 & 721 & 0.25 & 0.31 &  95.9 & 120.4 & 129.0 & 11.4 \\  
 15 & 1.49 & 4.49 & 5.4 & 0.33 & 1.1 & 0.246 & 0.308 & 841 & 783 & 745 & 0.16 & 0.27 &  97.6 & 115.7 & 130.2 & 12.6 \\  
 16 & 1.55 & 2.77 & 5.6 & 0.02 & 1.4 & 0.138 & 0.137 & 600 & 551 & 549 & 0.18 & 0.22 & 106.3 & 124.6 & 128.9 &  7.5 \\  
 17 & 1.70 & 3.06 & 5.0 & 0.03 & 1.3 & 0.175 & 0.213 & 646 & 591 & 582 & 0.16 & 0.22 & 104.4 & 122.3 & 129.4 &  9.4 \\ 
 18 & 1.76 & 3.49 & 5.3 & 0.13 & 1.2 & 0.233 & 0.303 & 744 & 685 & 658 & 0.18 & 0.29 & 100.2 & 119.1 & 131.5 & 12.1 \\  
 19 & 1.75 & 4.00 & 3.6 & 0.09 & 1.6 & 0.060 & 0.165 & 700 & 715 & 695 & 0.05 & 0.15 & 109.0 & 111.4 & 125.1 &  7.7 \\  
 20 & 1.76 & 4.50 & 4.2 & 0.14 & 1.2 & 0.109 & 0.290 & 690 & 685 & 656 & 0.06 & 0.19 & 104.9 & 111.0 & 129.9 & 13.2 \\  
 21 & 1.55 & 2.77 & 5.6 & 0.02 & 1.4 & 0.138 & 0.137 & 600 & 551 & 549 & 0.18 & 0.22 & 106.3 & 124.6 & 128.9 &  7.5 \\  
 22 & 1.76 & 3.18 & 5.1 & 0.05 & 1.4 & 0.184 & 0.274 & 686 & 623 & 613 & 0.19 & 0.28 & 103.9 & 123.1 & 133.2 & 11.3 \\ 
 23 & 1.95 & 3.59 & 4.6 & 0.06 & 1.5 & 0.138 & 0.239 & 644 & 620 & 599 & 0.11 & 0.21 & 105.3 & 116.2 & 128.9 & 10.4 \\  
 24 & 2.01 & 4.01 & 3.6 & 0.07 & 1.5 & 0.135 & 0.224 & 645 & 627 & 607 & 0.10 & 0.17 & 108.2 & 117.4 & 129.2 &  9.5 \\  
 25 & 1.99 & 4.52 & 3.8 & 0.15 & 1.4 & 0.006 & 0.051 & 589 & 615 & 635 & 0.03 & 0.05 & 111.2 & 110.8 & 112.5 &  3.0 \\
\hline
\label{tab:sims}
\end{tabular}
\end{center}
\end{minipage}
\end{table}
\end{landscape}

\section{Simulations}
\label{sec:sim}

We use the set of 25 simulations from \citet[][hereafter
W08]{Widrow2008}, which represents the evolution of a Milky Way-like
galaxy from idealized initial conditions.  The advantage of using
these simulations (aside from their high quality setup) is that they
provide a range of possible evolutionary paths for at least one
galaxy.  By restricting ourselves to models for a single galaxy we may
underestimate the expected scatter in the evolution.  Note that since
collisionless simulations can be rescaled in mass, size, and velocity
subject to the condition $G=1$, where $G$ is the gravitational
constant, our results can be applied to a fairly broad set of galaxy
mass.  Below we describe in brief the setup of the simulations and
refer the reader to W08 for a more detailed discussion.

\subsection{Galaxy models}

The initial conditions for the simulations are $N$-body realisations
of axisymmetric galaxy models that consist of a disc, a bulge and a
dark matter halo.  The distribution function for the composite system
is
\begin{eqnarray}
\label{eq:totalDF}
f\left( E,\,L_z,\,E_z \right)= f_{\rm d}\left( E,\,L_z,\,E_z \right)+f_{\rm b}\left( E\right) +f_{\rm h}\left(E\right),
\end{eqnarray}
where the energy $E$ and the angular momentum about the symmetry axis
$L_z$ are exact integrals of motion and $E_z$ is an approximate third
integral corresponding to the vertical energy of stars in the disc
\citep{Kuijken1995, Widrow2005}.  Since $E_z$ is very nearly conserved
for orbits that are not far from circular, the initial system will be
close to equilibrium so long as the disc is relatively ``cold'', a
condition met for the models considered in this paper.

The distribution function for the disc is constructed to yield the
density distribution \citep{Kuijken1995}
\begin{eqnarray}
\rho_{\rm d}\left( R,z\right) = \frac{M_{\rm d}}{2\pi R_{\rm
    d}^2}\,
e^{-R/R_{\rm d}} \,
{\rm sech}^2{\left (z/z_{\rm d}\right )}\,{\rm erfc}\left
  (\frac{r-R_t}{2^{1/2}\delta R_t}\right )
\end{eqnarray}
where $R$ and $z$ are cylindrical coordinates, $r$ the spherical
radius, $R_{\rm d}$ the scale-length, $z_{\rm d}$ the scale-height and
$M_{\rm d}$ the total mass of the disc.  The disc is truncated at
radius $R_t=10 R_{\rm d}$ with a truncation sharpness of $\delta
R_t=1$~kpc. The distribution function is constructed so that the
radial dispersion profile is exponential $\sigma_R^2(R) =
\sigma_{R0}^2 \exp{\left (-R/R_{\rm d}\right)}$ where $\sigma_R$ is
the radial velocity dispersion in cylindrical coordinates.

The bulge and halo distribution functions are designed so that their
respective density profiles approximate the user-specified functions
$\tilde{\rho}_{\rm b}$ and $\tilde{\rho}_{\rm h}$.  For the bulge, we
assume a ``target'' density profile
\begin{equation}
\tilde{\rho}_{\rm b}(r) = \rho_b\left (\frac{r}{R_e}\right )^{-p} e^{-b\left (r/R_e\right )^{1/n}}~,
\end{equation}
which yields, on projection, the S\'{e}rsic law with $p=1-0.6097/n +
0.05563/n^2$ \citep{Prugniel1997, Terzic2005} where $n$ is the
S\'{e}rsic index and $\rho_b$ is the central surface density. The
constant $b$ is adjusted so that \re\ contains half of the total
projected mass of the bulge. These models use
\begin{equation}\label{vbdef}
\sigma_b\equiv
\left (4\pi n b^{n(p-2)} \Gamma\left (n\left (2-p\right )\right )R_e^2 \rho_b \right )^{1/2}
\end{equation}
rather than $\rho_b$ to parametrise the overall density scale of the
bulge models, where $\Gamma$ is the gamma function.

The target halo density profile is
\begin{equation}
\tilde{\rho}_{\rm halo}(r) = \frac{2^{2-\gamma}\sigma_h^2}{4\pi a_h^2}\frac{1}{\left (r/a_h\right )^{\gamma}\left (1 + r/a_h\right )^{3-\gamma}}\,C\left (r;r_h,\delta r_h\right )~,
\end{equation}
where $\gamma=1$ is the central cusp strength, $a_h$ is the
scale-length and $C\left (r;r_h,\delta r_h\right )$ is a truncation
function that decreases smoothly from unity to zero at $r\simeq r_h$
within a radial range $\delta r_h$.  The models considered here assume
$r_h=100$~kpc and $\delta r_h=5$~kpc and use the function $C\left
  (r;r_h,\delta r_h\right ) = \frac{1}{2} {\rm erfc}\left
  (\left(r-r_h\right)/\sqrt{2}\delta r_h\right )$.

The bulge and halo distribution functions, which, by assumption,
depend only on the energy, are found via an inverse Abel transform
\citep{BinneyTremaine}.  Since this method assumes spherical symmetry
we first calculate an approximate spherically-averaged total potential
\begin{equation}
\tilde{\Psi}_{tot} = 
\tilde{\Psi}_{\rm d} +
\tilde{\Psi}_{\rm b} + 
\tilde{\Psi}_{\rm h}  
\end{equation}
where $\tilde{\Psi}_{\rm d}$ is the monopole term of a spherical
harmonic expansion for the disc and $\tilde{\Psi}_{\rm b,h}$ are
calculated from $\tilde{\rho}_{\rm b,h}$.  We then evaluate
\begin{equation}\label{eq:abel}
f_{b,h}\left (E\right ) = \frac{1}{\sqrt{8}\pi^2}\int_E^0 \frac{d^2\tilde{\rho}_{b,h}}{d\tilde{\Psi}_{\rm tot}^2}\frac{d\tilde{\Psi}_{\rm tot}}{\sqrt{\tilde{\Psi}_{\rm tot}-E}}.
\end{equation}
Armed with the distribution functions for the three components we
solve Poisson's equation in {\it axisymmetry} using an iterative
scheme and Legendre polynomial expansion \citep{Kuijken1995,
  Widrow2005}.  Note that the bulge and halo are flattened slightly
due to the influence of the disc potential.

\subsection{Model parameters}

The models described above were tailored to satisfy observational
constraints for the Milky Way such as the inner and outer rotation
curve, the local vertical force, the line-of-sight velocity dispersion
toward Baade's window, and the circular speed at the position of the
Sun.  A Bayesian/MCMC algorithm provided the probability distribution
function (PDF) of Milky Way models over the model parameter space.
Models from the PDF span a wide range of structural properties.  For
example, $M_{\rm d}$ varies in the range $2-7\times10^{10}$ \Msun,
while $R_{\rm d}$ varies between $2.0$ kpc and $3.5$~kpc.

\begin{figure}
\centering
\includegraphics[width=0.6\hsize,angle=270]{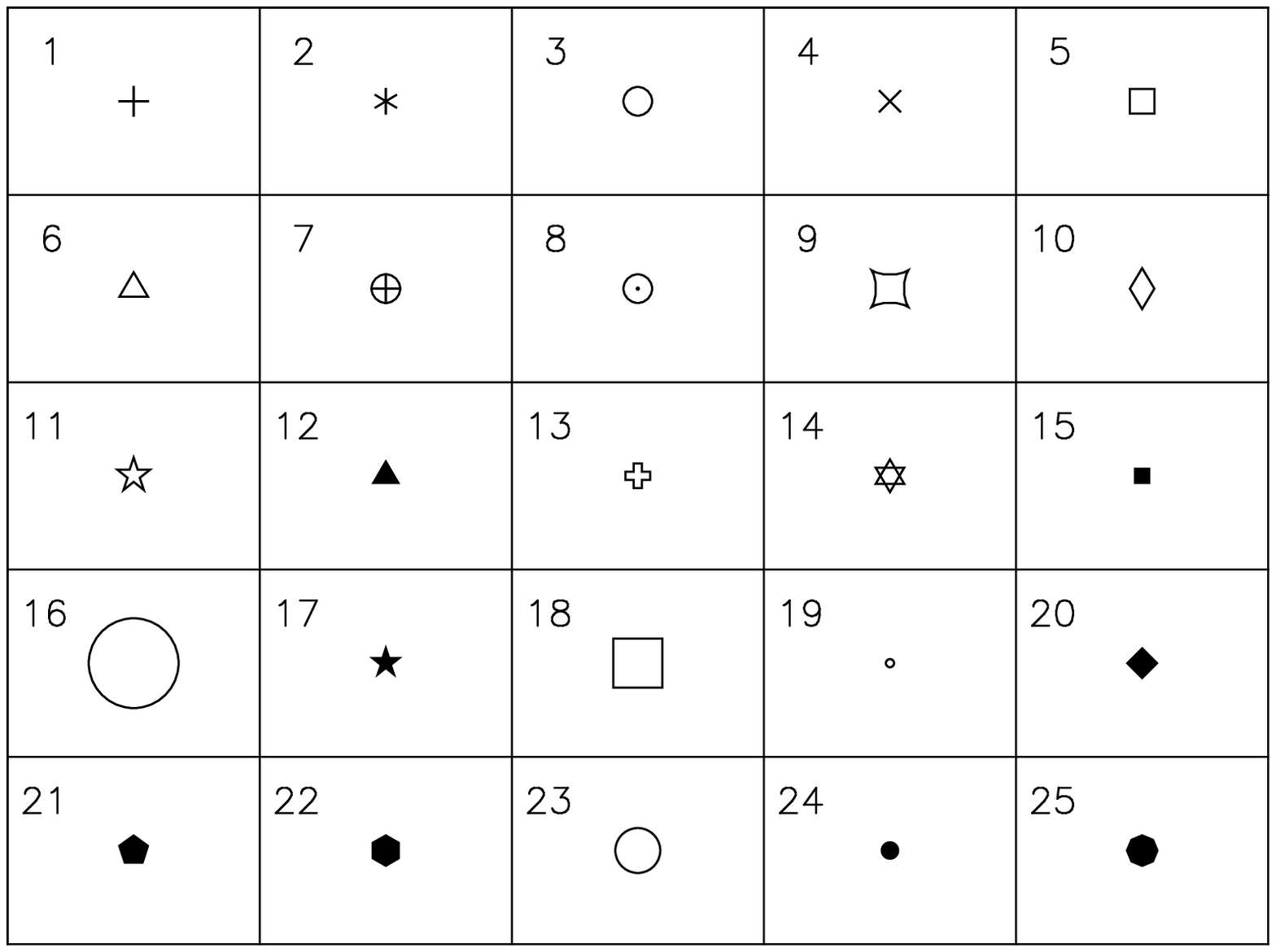} 
\caption{A grid of identifying symbols used to represent each model in
  the various plots of this paper.}
\label{fig:symbols}
\end{figure}

The stability of a stellar disc is determined by two parameters:
\begin{eqnarray}
Q=\frac{\sigma_R\kappa}{3.36G\Sigma} ~{\rm and}~X\equiv\frac{\kappa^2R}{2\pi G\Sigma m}~,
\end{eqnarray}
where $\kappa$ is the epicyclic radial frequency, $G$ is the
gravitational constant, $\Sigma$ is the surface density, $R$ is the
radius and $m$ is the azimuthal mode number of the perturbation
\citep{Toomre1981, Toomre1964, Goldreich1978, Goldreich1979}. Here we
take $m=2$ since we are interested in bars.  We select 25 models that
span the region of the $Q-X$ plane where the PDF is non-negligible
($1.0\lesssim Q\lesssim 2.0$ and $2.5\lesssim X \lesssim 4.0$ (see
Section 7 of W08).  The properties of the models are summarized in
Table~\ref{tab:sims} while the symbols used to represent each model
are shown in Fig.~\ref{fig:symbols}.  For instance, a plus ('+')
symbol is used to represent model 1 throughout the paper.

The models do not contain a SMBH since an initial SMBH satisfying the
\Msige\ relation would have a mass of only $\sim 10^7$~\Msun. The
influence radius of this SMBH would be $G\Mbh/\sigma^2 \approx 10$~pc,
which is smaller than the softening length used.

\subsection{Numerical parameters}

The bulge, disc and halo consist of $2\times10^5$, $6\times10^5$ and
$1\times10^6$ particles respectively.  The particle softening
$\epsilon=25$~pc for all particles and the models were evolved for
$10^4$ equal time steps of length $\Delta t=0.5$~Myr.  The 25 models
were evolved for 5~Gyr using the parallel $N$-body tree code described
in \citet{Dubinski1996}.


\section{Evolution of central density and velocity dispersion}
\label{sec:bar}

\begin{figure}
\centering
\includegraphics[width=0.8\hsize,angle=270]{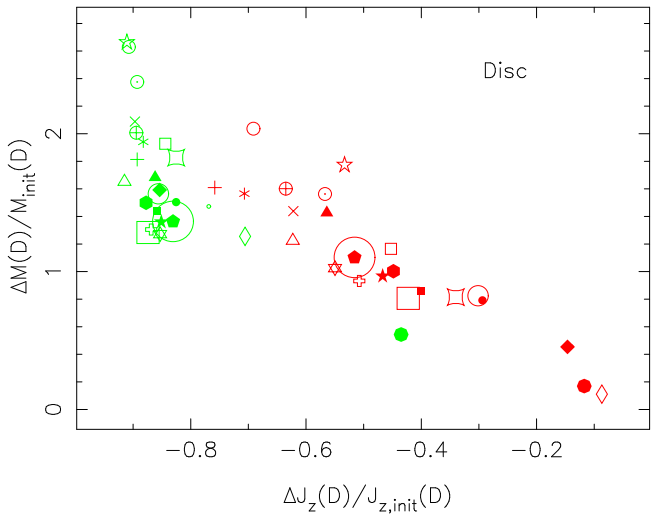} 
\caption{The fractional change in mass of the disc within the bulge
  \re\ (at $t_2$) plotted versus the changes in total angular momentum
  at $t_1$ in red and at $t_2$ in green.}
\label{fig:angmom}
\end{figure}

All the models formed bars. We measure the bar amplitude, \Abar, as
the normalised amplitude of the $m=2$ Fourier moment of the surface
density of disc particles:
\begin{eqnarray}
\Abar=N_{d}^{-1}\left|\sum_{j~\in~disc}e^{2i\phi_j}\right|~, 
\end{eqnarray}
where $\phi_j$ is the two-dimensional cylindrical polar angle in the
equatorial plane of the disc for the $j$th particle, and $N_d$ is the
total number of disc particles.  We consider three different times in
the simulations $t_0=0$, $t_1=2.5$~Gyr and $t_2=5$~Gyr and refer to
these times throughout the paper. In most simulations the bar forms by
1~Gyr and continues to grow until $t_2$ (see Fig.~17 in W08), while
in simulations 1 and 12 the bar amplitude peaks at 0.5 and 2~Gyr,
respectively, and declines slightly thereafter.  Values of \Abar\ at
$t_1$ and $t_2$ are given in Table~\ref{tab:sims}.

We obtain the bulge \re\ by calculating the face-on projected circular
aperture containing half of the bulge particles. We also measured \re\
by fitting a S\'ersic profile to the mass-weighted surface density
profile and found consistent values.  We find \re\ in the range $456$
pc $<\re<841$~pc at $t_0$, decreasing to $439$ pc $<\re<747$~pc at
$t_2$, except in simulation 25, where \re\ increases slightly (see
Table~\ref{tab:sims}).  Throughout the paper all measurements,
including those for the disc (D) and bulge$+$disc (B+D), are computed
within \re\ {\it of the bulge} (B).

The formation and growth of a bar leads to the outward transport of
angular momentum \citep{Debattista2000, Athanassoula2002} resulting in
an increase in the mass fraction of the disc in the central region, as
was originally shown by \citet{Hohl1971}.  We quantify the fractional
change in the mass of the central region by defining $\Delta {\rm
  M/M_{init} = \left(M_{t} - M_{init}\right)/M_{init}}$, where ${\rm
  M_{t}}$ is the mass within \re\ at either $t_1$ or $t_2$, and ${\rm
  M_{init}}$ is the mass within \re\ at $t_0$.
The contribution of the halo mass within $r < \re$ is less than $25\%$
of the total mass; we therefore neglect the dark matter particles in
our analysis.  We measure the change in angular momentum by defining
$\Delta J_z\D/J_{z,{\rm init}}\D = \left(J_{z,t}\D - J_{z,{\rm
      init}}\D \right)/J_{z,{\rm init}}\D$, where $J_{z,t}\D$ is the
angular momentum at $t_1$ or $t_2$ of disc particles within \re\ and
$J_{z,{\rm init}}\D$ is the angular momentum at $t_0$ of all disc
particles within \re.  We use \re\ for bulge particles measured at
$t_2$ in order that the changes plotted are due to a difference in
angular momentum, rather than different radial range.  In Fig.
\ref{fig:angmom} we show that the fractional change of the total
angular momentum leads to an increase in the central mass of the disc
and that the change in angular momentum reaches $\sim -90\%$ by $t_2$.

\begin{figure*}
\centering
\begin{tabular}{ccc}
\includegraphics[width=0.37\hsize,angle=270]{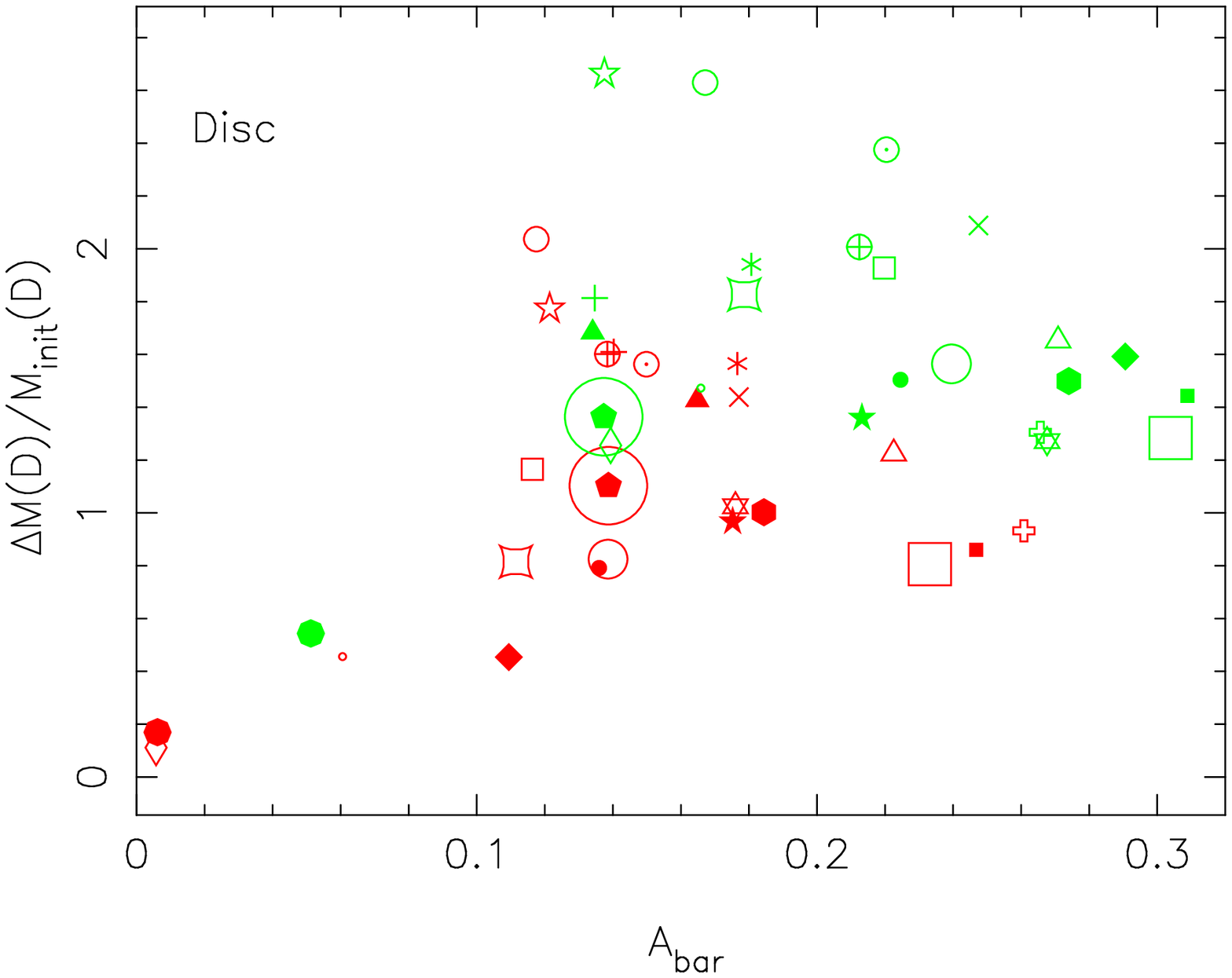} & 
\includegraphics[width=0.37\hsize,angle=270]{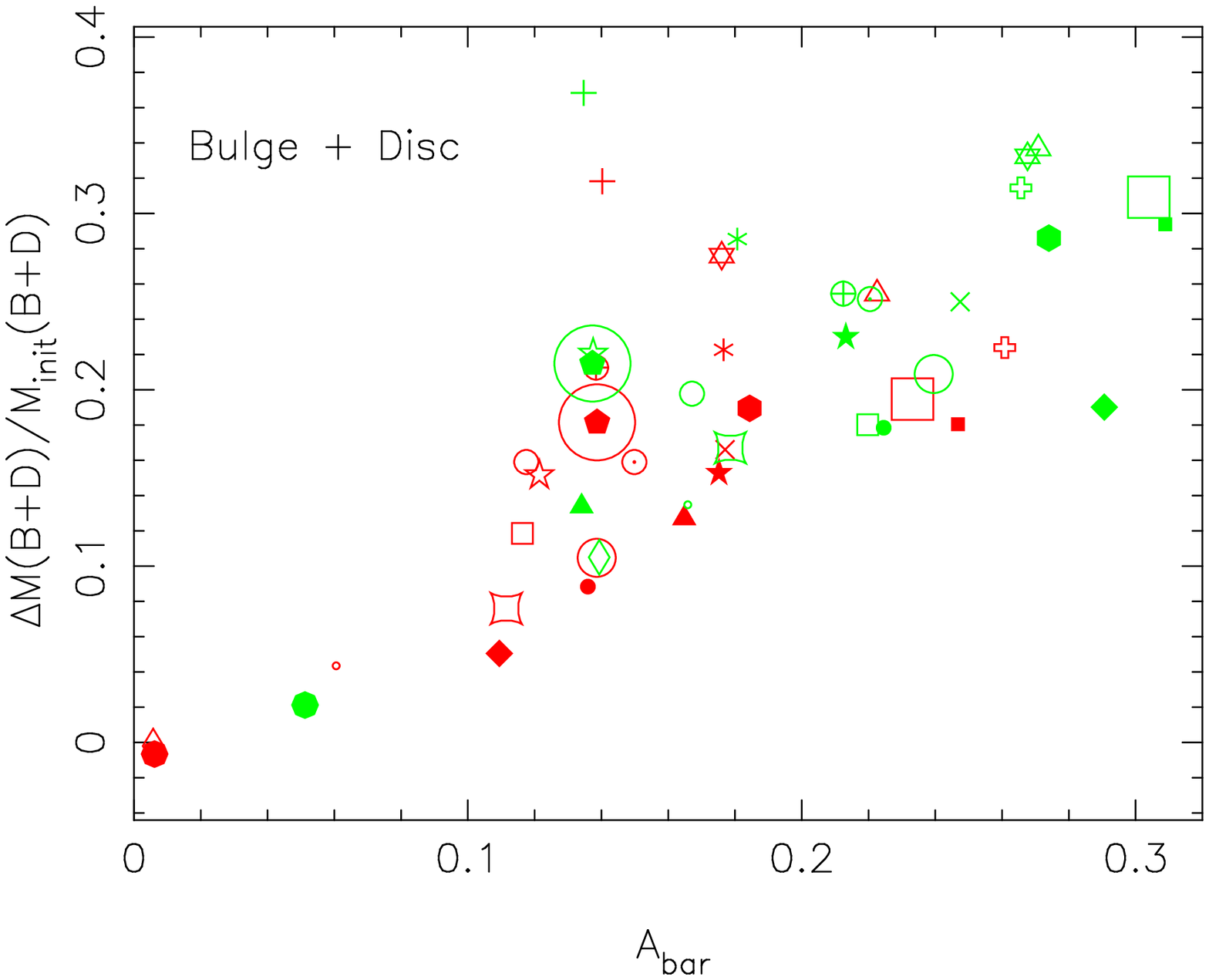} \\
\includegraphics[width=0.37\hsize,angle=270]{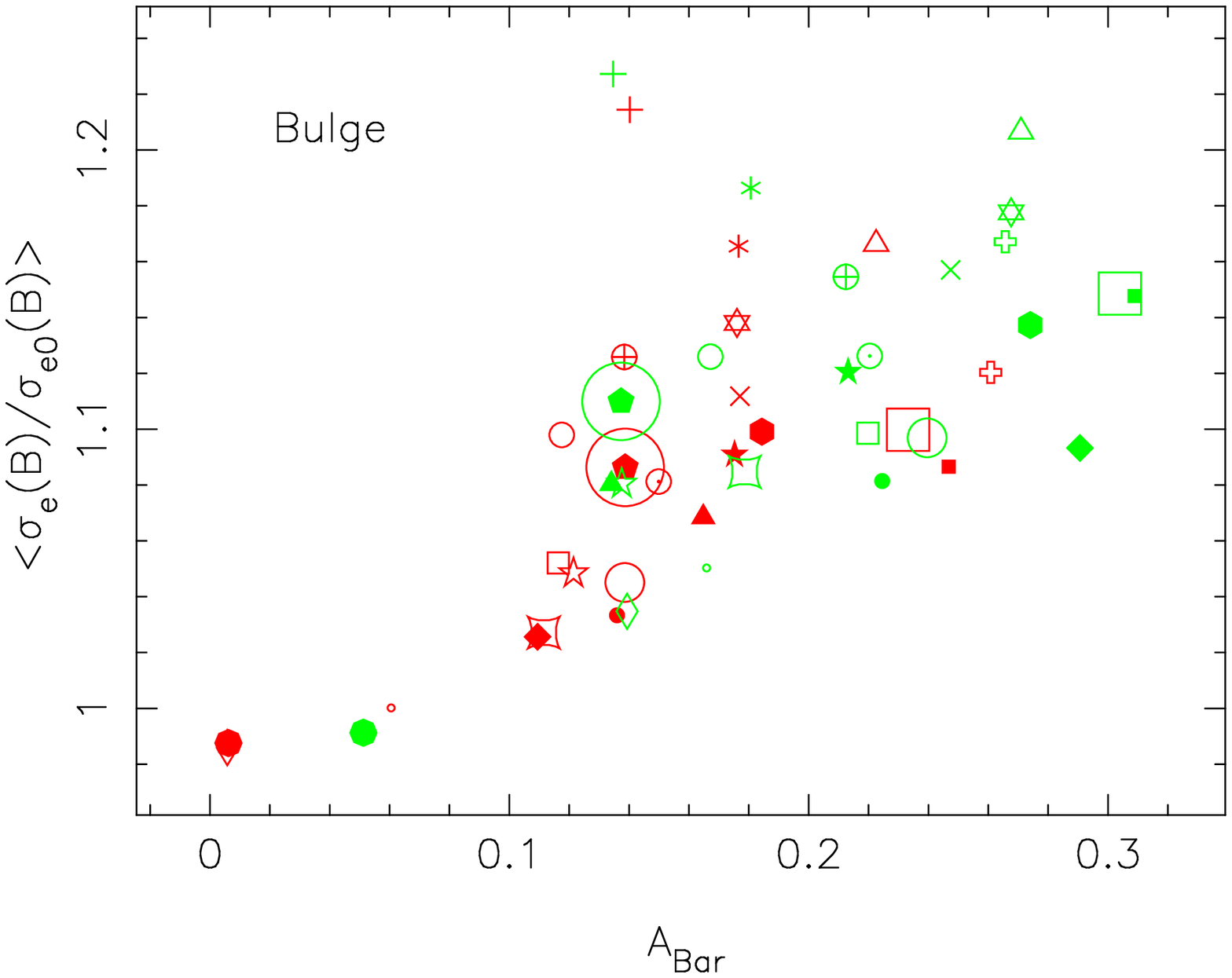} &
\includegraphics[width=0.37\hsize,angle=270]{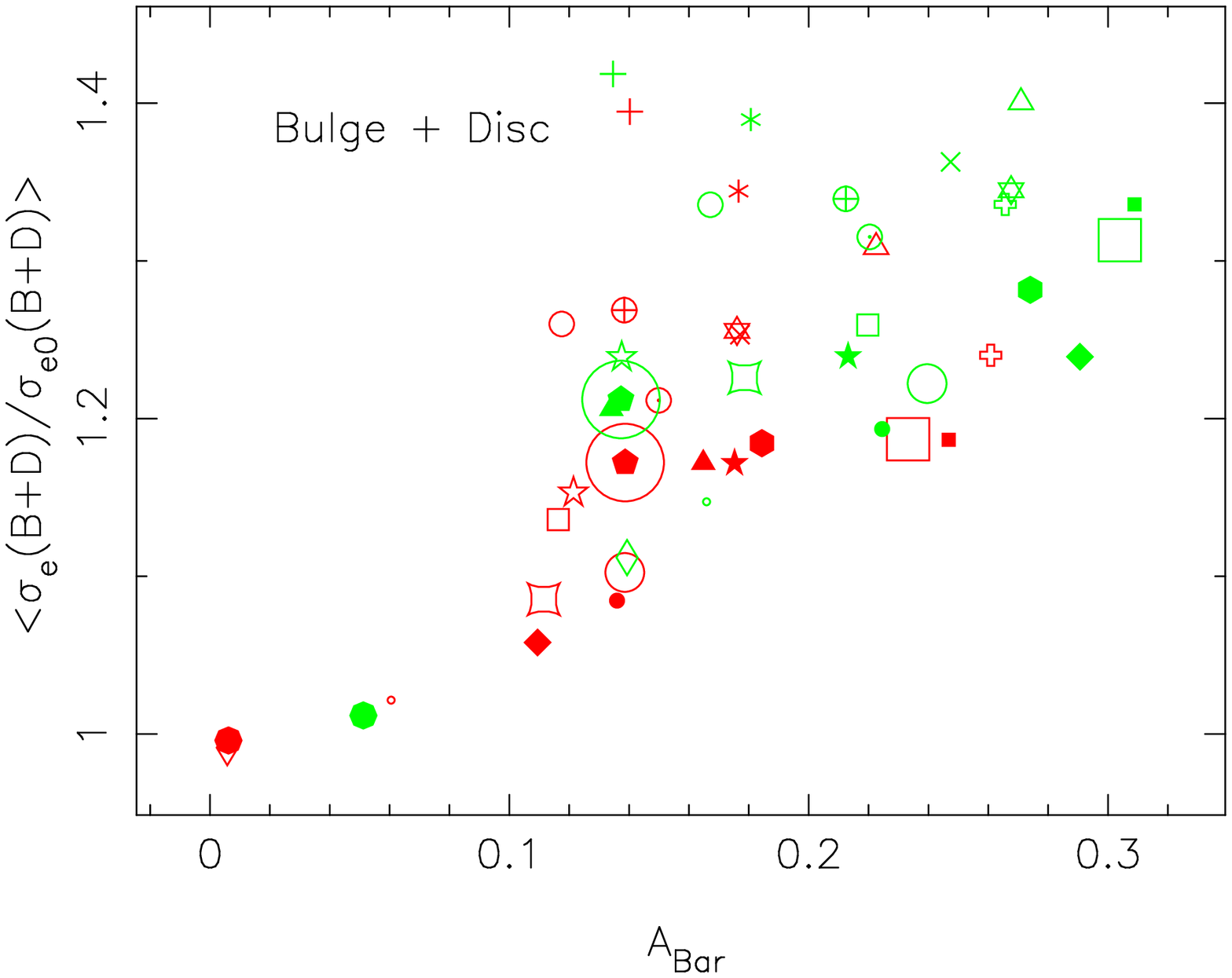} 
\end{tabular}
\caption{Top row: The fractional changes in the mass of the disc (left
  panel) and bulge$+$disc (right panel) within \re\ of the bulge
  plotted versus the bar amplitude \Abar. Bottom row: The average
  ratio of final to initial velocity dispersion,
  $\left<\sige/\sigma_{e0}\right>$, for bulge particles (left panel)
  and for bulge$+$disc particles (right panel) versus \Abar. In all
  panels values at $t_1$ are shown in red, and at $t_2$ in green. }
\label{fig:bar}
\end{figure*}

In Fig.~\ref{fig:bar} we plot the fractional change in mass, $\Delta
{\rm M/M_{init}}$, versus the bar strength \Abar.  The increase in
mass of the disc particles (Fig.~\ref{fig:bar} top-left) shows a large
scatter, with many models increasing by a factor of $1.5$ to $2$ by
$t_2$.  The top-right panel shows that the fractional mass increase
for disc$+$bulge particles instead correlates with bar strength.  The
bulge-to-disc mass ratio within \re\ is $2.8\lesssim
B/D\left(R<\re\right)\lesssim8.8$ initially, decreasing to
$1.2\lesssim B/D\left(R<\re\right)\lesssim7.7$ by $t_2$.

\subsection{Measuring velocity dispersions}

The increase in the central density deepens the potential and raises
the velocity dispersion \sige\ \citep{Debattista2013}.  We define
\sige\ as the mass-weighted aperture velocity dispersion within a
circular aperture of radius \re:
\begin{eqnarray}
 \sige^2 & = & \frac{\int^{R_e}_0 I\left(R\right)\left(\sigma_{los}^2\left(R\right)+\bar{v}_{los}^2\left(R\right)\right)dR}{\int^{R_e}_0 I \left(R\right)dR}
\end{eqnarray}
where $I \left(R\right)$ is the mass density, $\sigma_{los}$ is the
standard deviation and $\bar{v}_{los}$ is the mean line-of-sight
velocity of particles within \re.  For a particle distribution this
becomes:
\begin{eqnarray}
 \sige^2 & = & \frac{\sum_{r_i\leq R_e} m_i v_{i,los}^2}{\sum_{r_i\leq R_e} 
m_i} 
\end{eqnarray}
where $r_i$ is the radius, $m_i$ is the mass and $v_{i,los}$ is the
line-of-sight velocity of the $i$th particle and the sum is over all
particles within the circular aperture.

For each model we measure \sige\ for four different bar position
angles ${\rm PA}=0\degrees$ (bar seen side-on), $30\degrees$,
$60\degrees$ and $90\degrees$ at four inclinations $i=0\degrees$
(face-on), $30\degrees$, $60\degrees$, $90\degrees$ (edge-on). We
define $\left< \sige\right>$ as the average of \sige\ measured over
the various orientations. The standard deviation of \sige\ over all
viewing angles is defined as the scatter $\Delta\sige$.  In
Fig.~\ref{fig:bar} (bottom row) we plot the average ratio of final
to initial velocity dispersion, $\left< \sige/\sigma_{e0}\right>$,
versus \Abar, where $\sigma_{e0}$ is \sige\ at $t_0$. Generally \sige\
increases with increasing bar strength, with \sige\BD\ increasing by
as much as $\sim40\%$.

\begin{figure*}
\centering
\begin{tabular}{cc}
\includegraphics[width=0.3\hsize,angle=270]{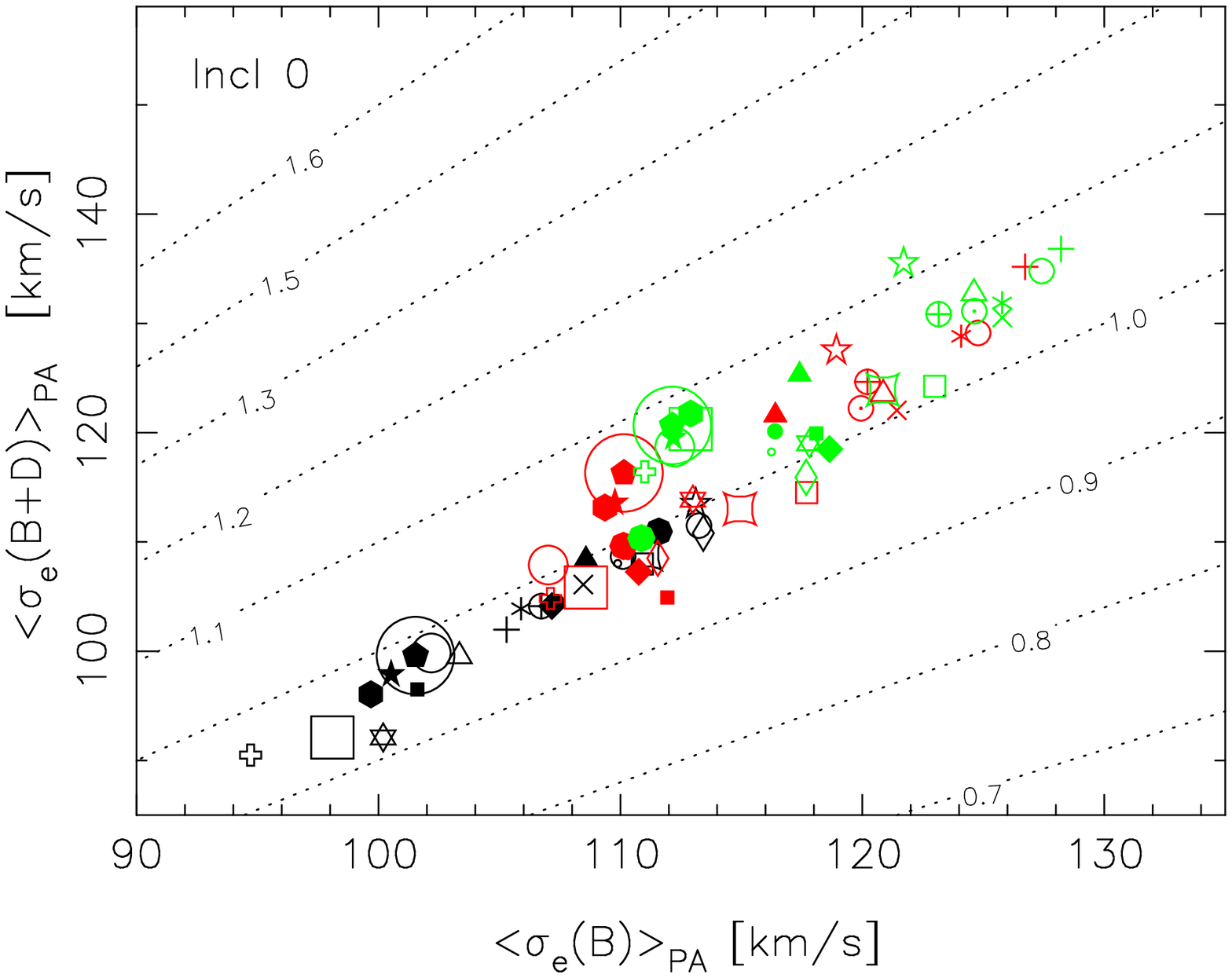} &
\includegraphics[width=0.3\hsize,angle=270]{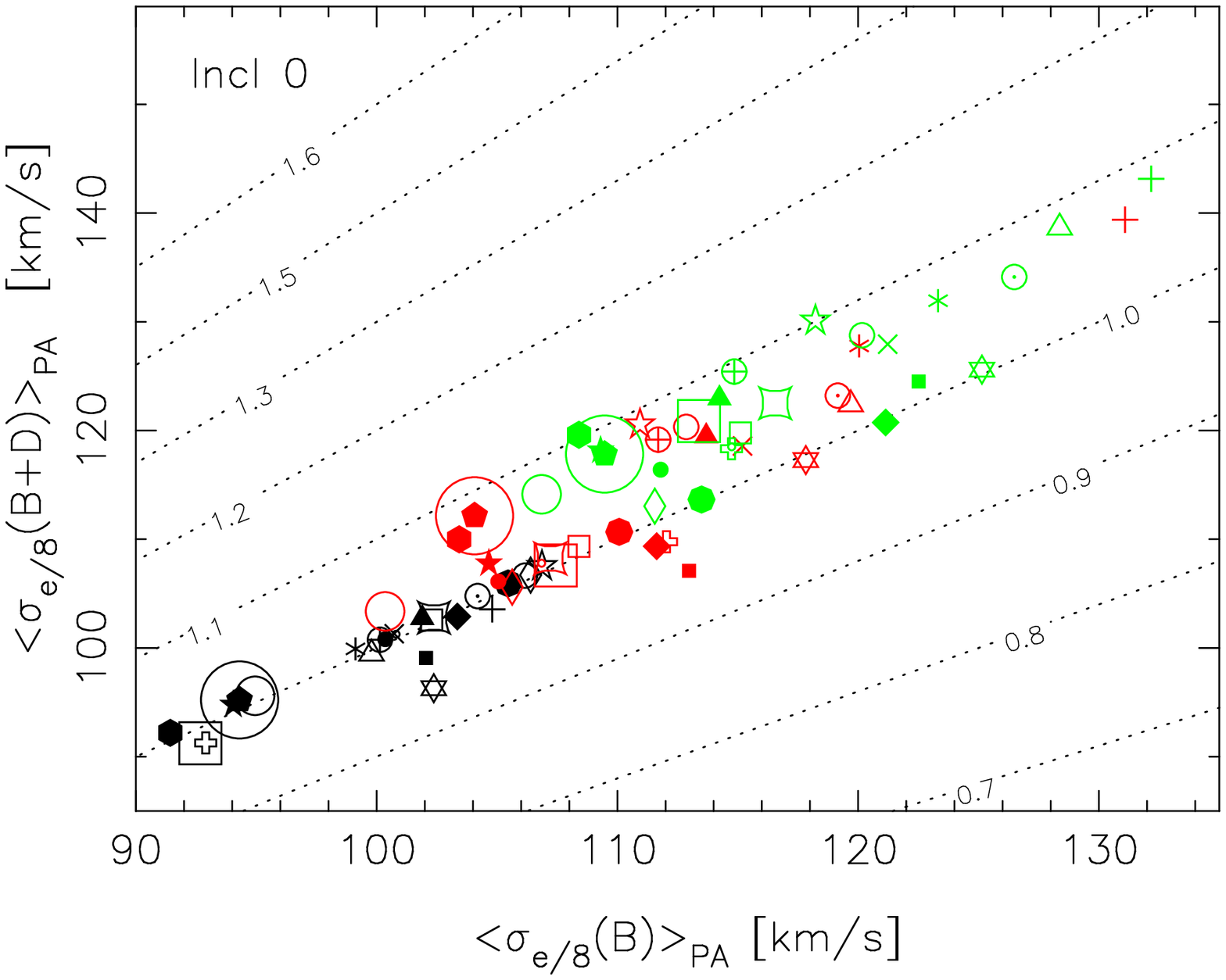}\\
\includegraphics[width=0.3\hsize,angle=270]{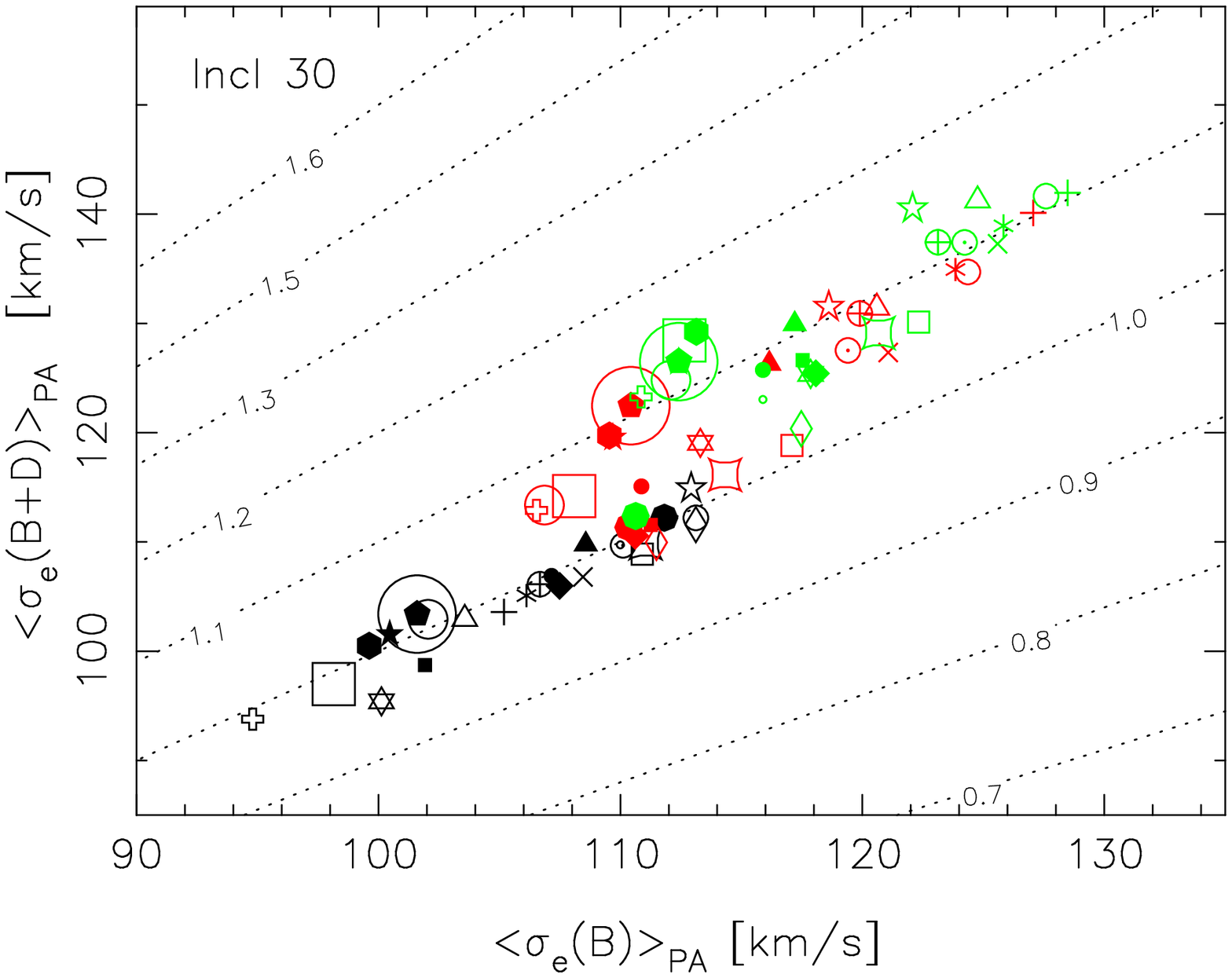} &
\includegraphics[width=0.3\hsize,angle=270]{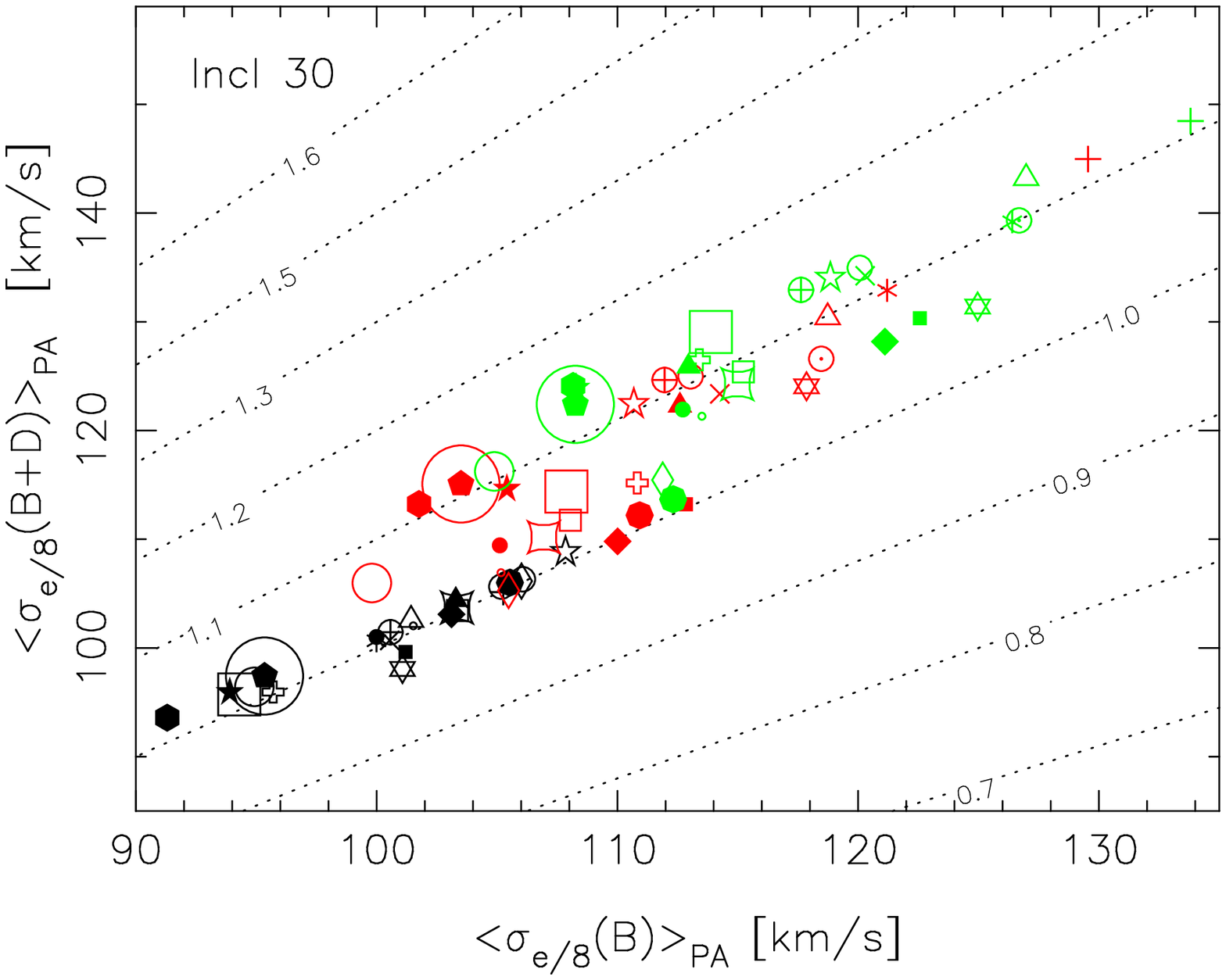} \\
\includegraphics[width=0.3\hsize,angle=270]{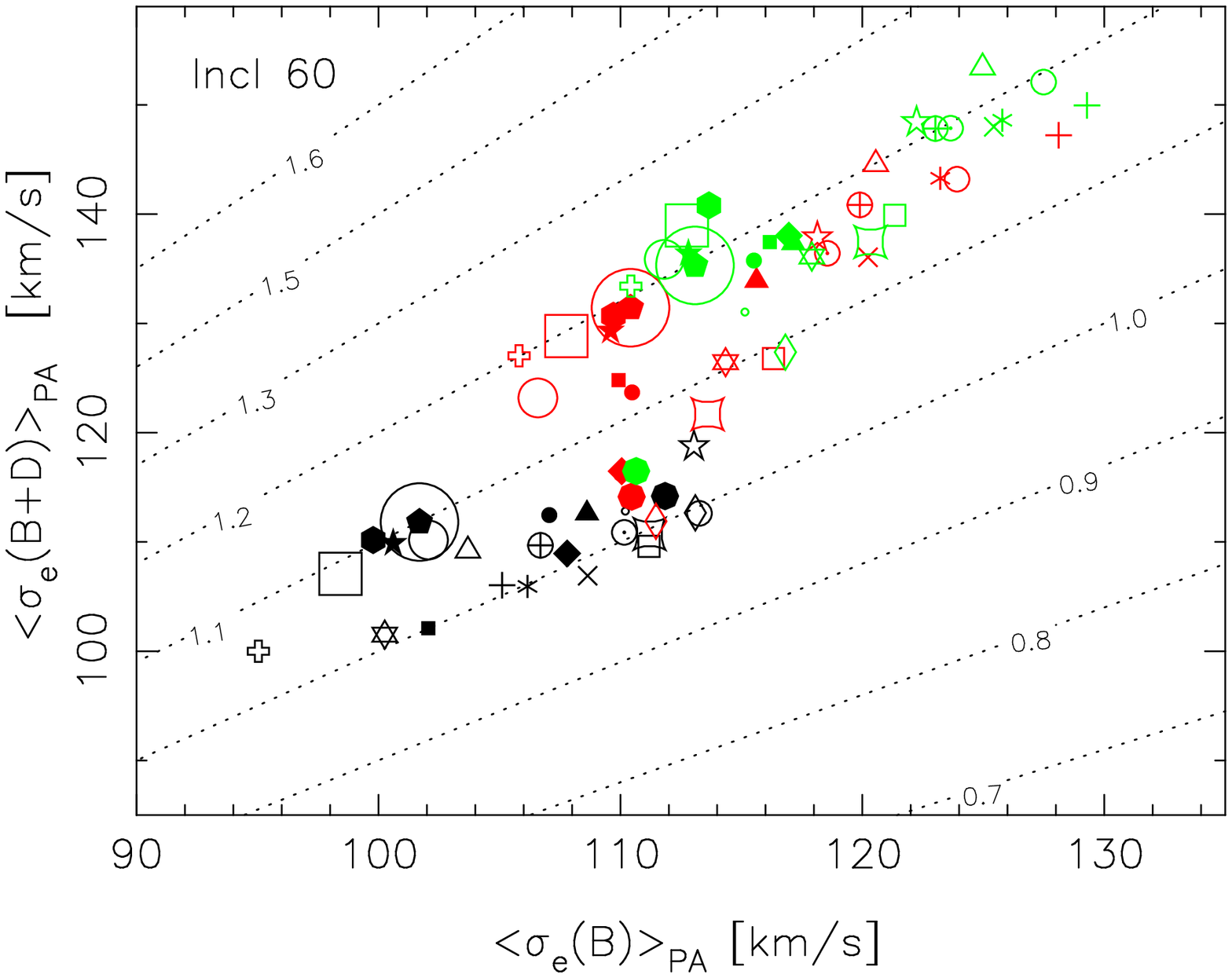} &
\includegraphics[width=0.3\hsize,angle=270]{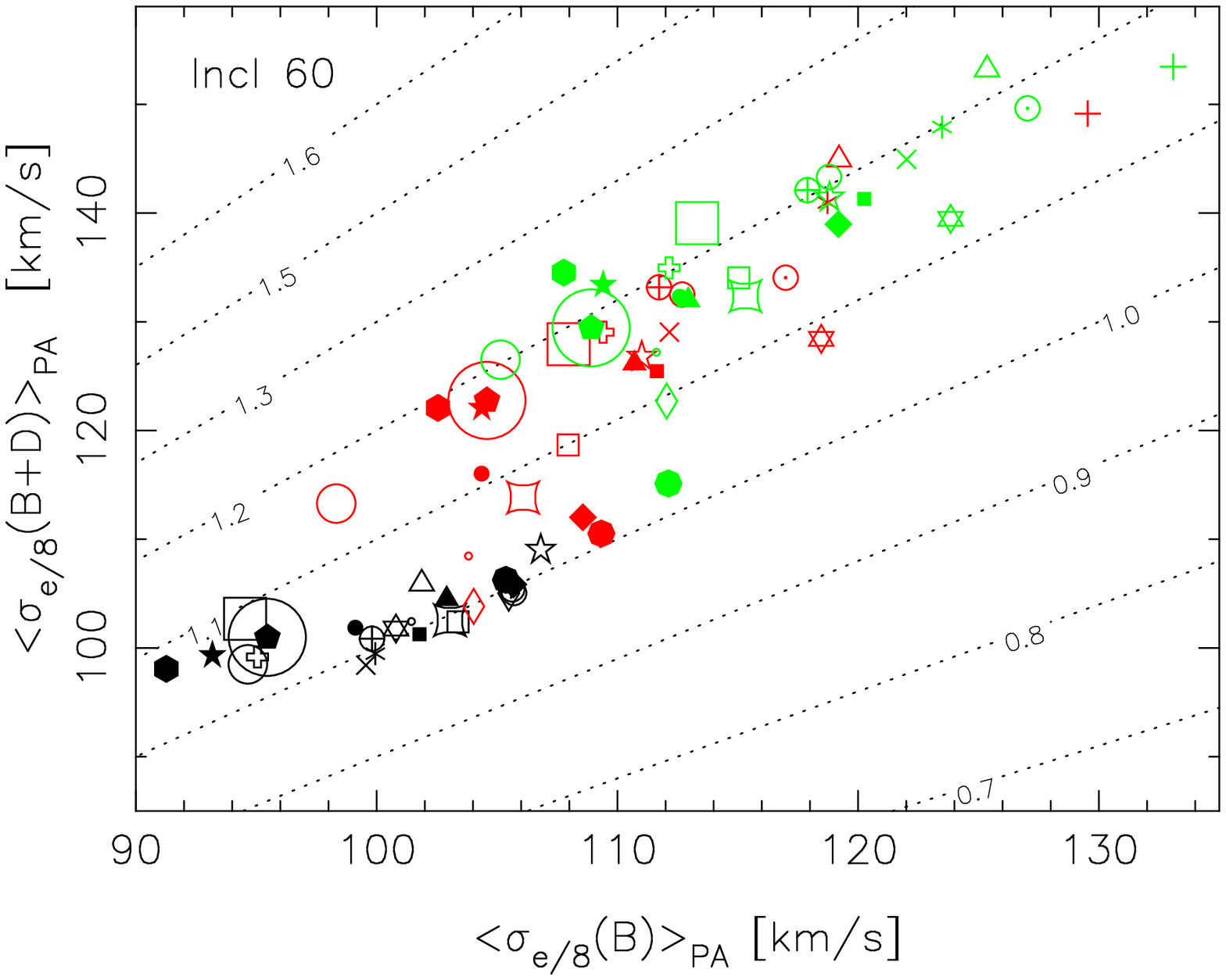} \\
\includegraphics[width=0.3\hsize,angle=270]{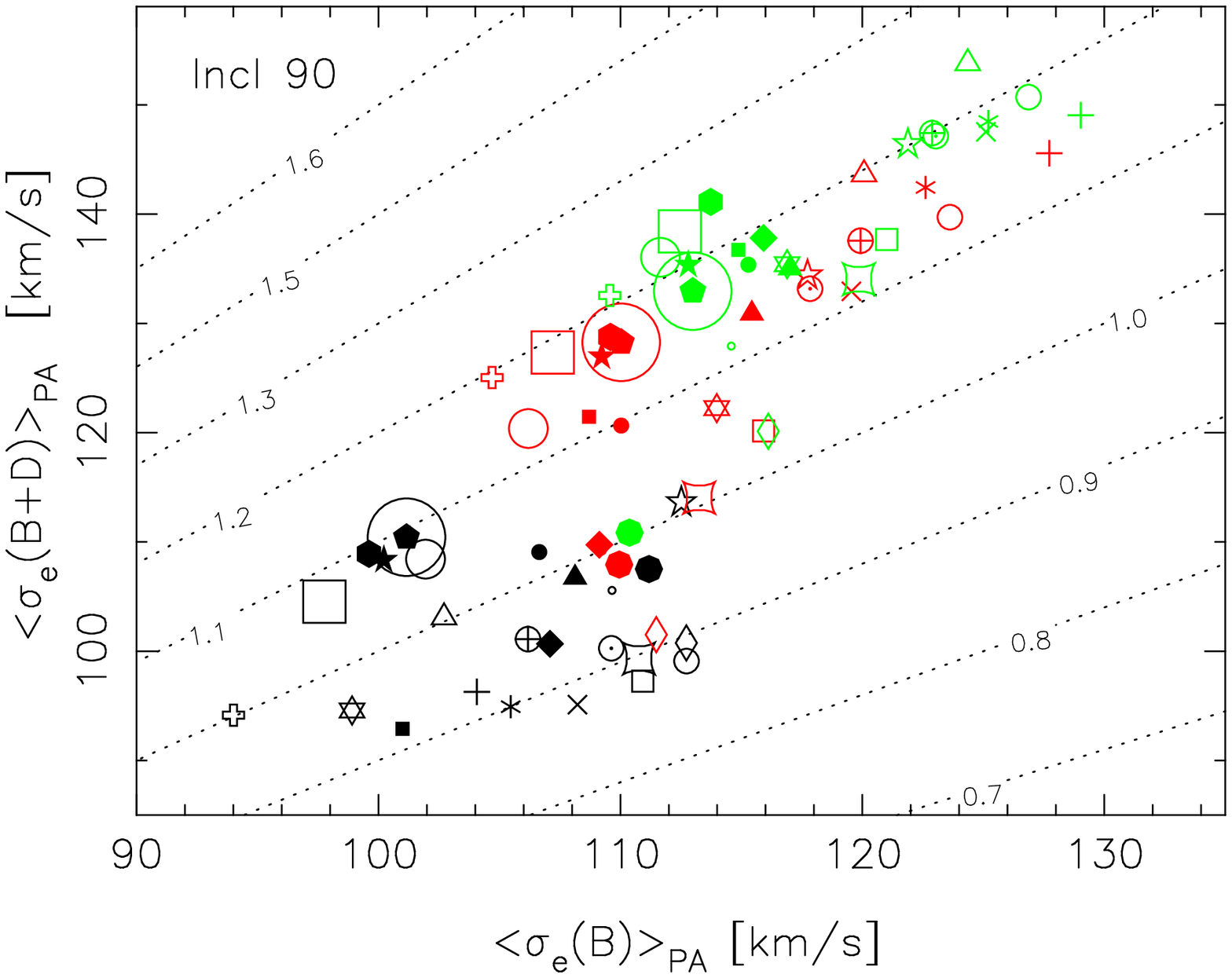} &
\includegraphics[width=0.3\hsize,angle=270]{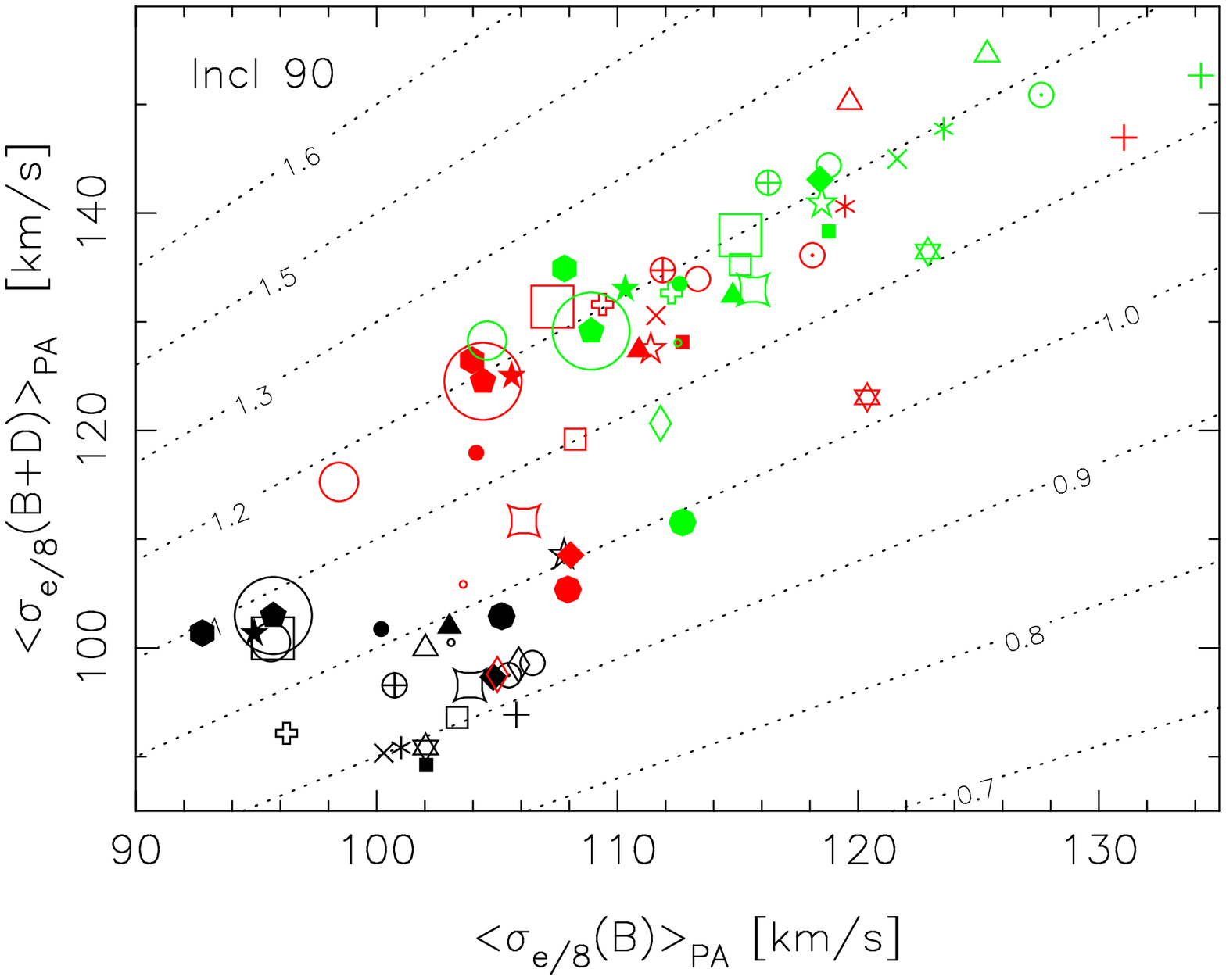}
\end{tabular}
\caption{Mean \sige\BD\ versus mean \sige\B\ (left panels) and mean
  \sigeight\BD\ versus mean \sigeight\B\ (right panels).  Black, red
  and green points represent the models at $t_0$, $t_1$ and $t_2$,
  respectively.  We average over PA$=0\degrees,30\degrees,60\degrees$
  and $90\degrees$ and plot $\sigma$'s for inclinations
  $i=0\degrees,30\degrees,60\degrees$ and $90\degrees$ from top to
  bottom.  Dotted lines have constant slope, as indicated along each
  line. }
\label{fig:correlation}
\end{figure*}

\subsection{The effect of disc contamination on velocity dispersions}
\label{sec:disc}

Assuming that the fundamental parameter which determines \Mbh\ is
\sige\ of the bulge only, disc contamination of \sige\ measurements
can lead to offsets in the \Msige\ relation for any galaxy.  Naively,
one way of reducing this contamination might seem to be to use a
smaller aperture since the ratio of bulge-to-disc mass within a given
aperture generally increases as the aperture is made smaller.  For
instance, within $\re/8$, the initial $B/D$ of the models is
$3.3\lesssim B/D\left(R<\re/8\right) \lesssim 47.9$ becoming
$1.4\lesssim B/D \left(R<\re/8\right) \lesssim13.4$ at $t_2$, which
can be compared with the smaller values discussed above.  We therefore
test whether the effect of disc contamination to the dispersion can be
reduced by using \sigeight.

\begin{figure}
\centering
\begin{tabular}{cc}
\includegraphics[angle=270,width=0.46\textwidth,]{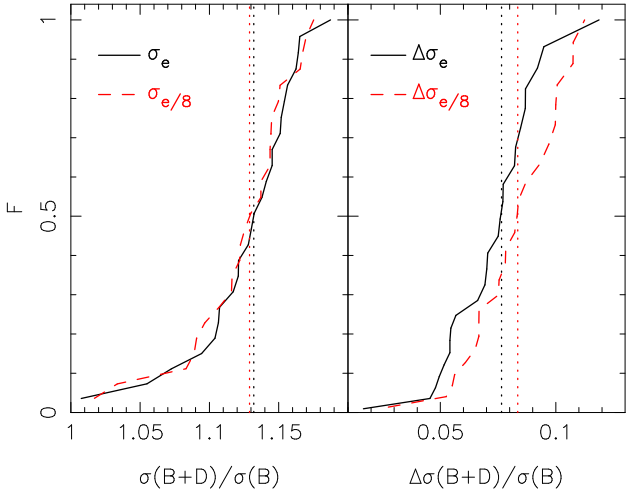}
\end{tabular}
\caption{Left panel: The cumulative distribution of
  \sigeight\BD/\sigeight\B\ (red dashed line) and \sige\BD/\sige\B\
  (black solid line).  Right panel: The cumulative distribution of
  $\Delta\sigeight\BD/\sigeight\B$ and $\Delta\sige\BD/\sige\B$.  The
  dotted lines show the median of each cumulative distribution. All
  distributions are shown at $t_2$.}
\label{fig:cumdist}
\end{figure}

In Fig.~\ref{fig:correlation} we compare \sige\B\ with \sige\BD\
within \re\ (left column) and \sigeight\B\ with \sigeight\BD\ within
$\re/8$ (right column) for four different inclinations.  In all cases
the general effect of disc contamination is to increase the
dispersion.  This is, on average, a $10\%$ effect in face-on galaxies
becoming $\sim 25\%$ for edge-on systems, in good agreement with
\citet{Debattista2013}.  This is true for both \sige\ and for
\sigeight.  Surprisingly, the effect of disc contamination on
\sigeight\ is about the same as on \sige.  In Fig.~\ref{fig:cumdist}
we plot the cumulative distribution of $\sigeight\left({\rm B+D}
\right)/\sigeight\left({\rm B}\right)$ and of $\sige\left({\rm B+D}
\right)/\sige\left({\rm B}\right)$. The two distributions are very
similar and the median of both distributions is $\sim1.13$.  A
Kolmogorov-Smirnov (K-S) test shows that the probability that the two
distributions are identical is $0.88$ showing that the aperture within
which the velocity dispersion is measured has little effect on
reducing the contamination from the disc.  We also plot the
distributions of $\Delta\sigeight\left({\rm B+D}
\right)/\sigeight\left({\rm B}\right)$ and of $\Delta\sige\left({\rm
    B+D} \right)/\sige\left({\rm B}\right)$, which show that the
scatter in \sigeight\BD\ is slightly larger than in \sige\BD: the
median of $\Delta\sigeight\BD/\sigeight\B$ is $0.084$ while for
$\Delta\sige\BD/\sige\B$ it is $0.077$.  The K-S test now finds that
the probability that both distributions are identical is only $0.41$.

We conclude that \sigeight\ does not provide any notable reduction in
the amount of contamination by the disc, while increasing slightly the
scatter in the measured dispersion.  Moreover, smaller apertures are
more likely to be contaminated by other nuclear components
\citep[e.g.][]{McConnell2013}.


\subsection{The effect of angular momentum redistribution on velocity
  dispersions}
\label{sec:anisotropy}

\begin{figure}
\includegraphics[width=0.8\hsize,angle=270]{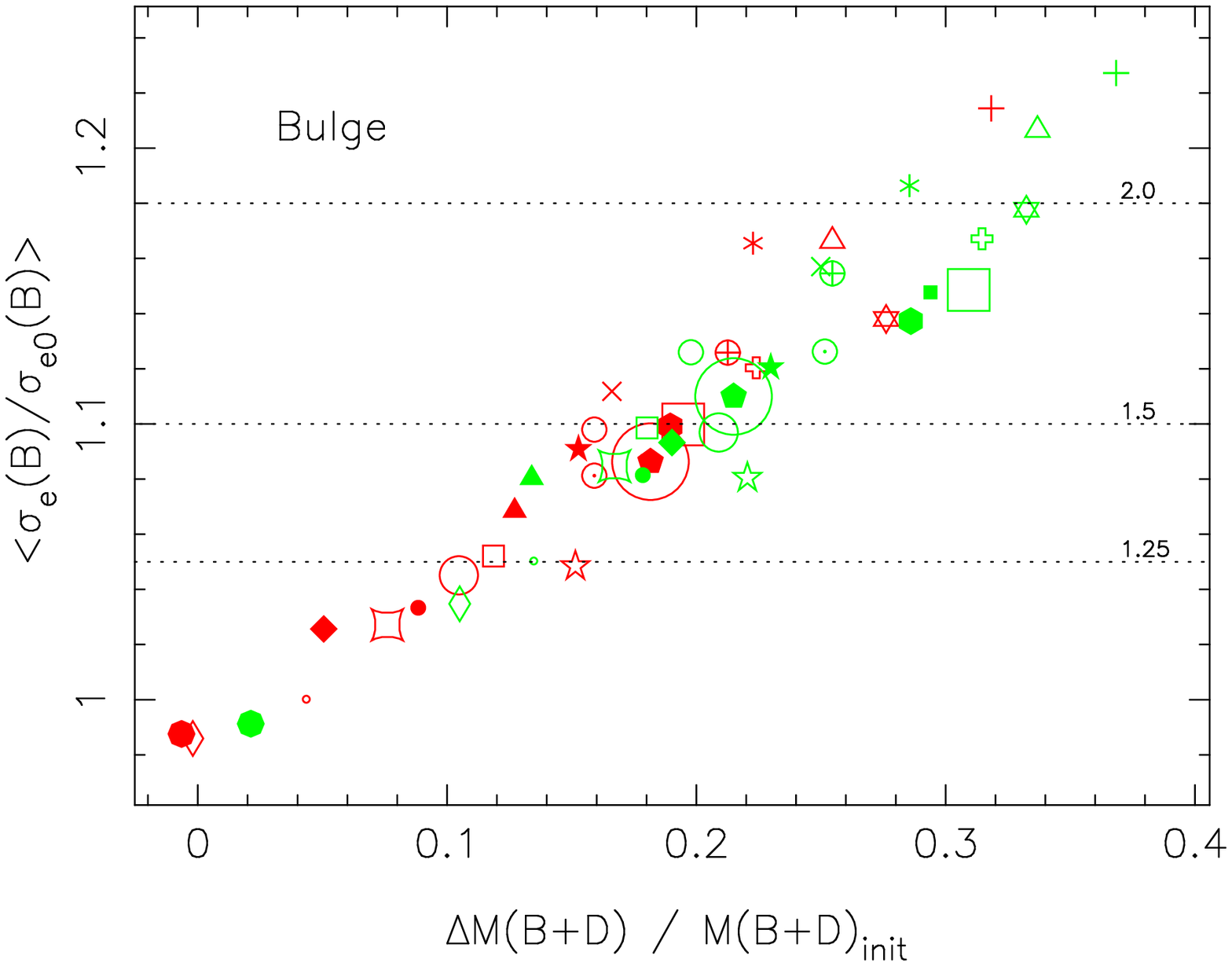} 
\includegraphics[width=0.8\hsize,angle=270]{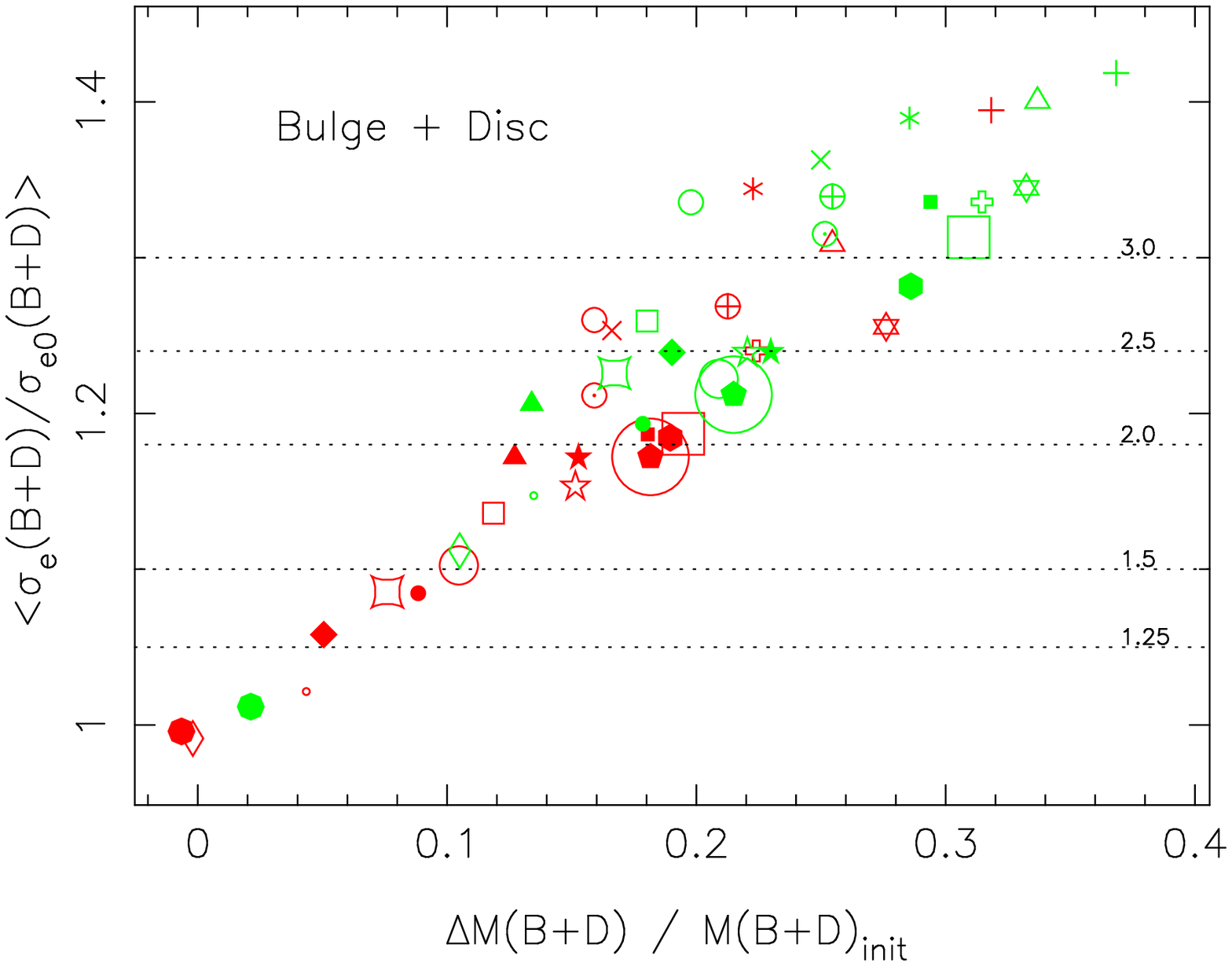} 
\caption{Average ratio of final to initial velocity dispersion at
  $t_1$ (red) and $t_2$ (green) for the bulge (top) and for the
  bulge$+$disc (bottom) versus the fractional change in mass of the
  bulge$+$disc within \re.  In both panels the dotted lines indicate
  contours of constant $\left(\sige/\sigma_{e0}\right)^\beta$ for
  $\beta=4.24$, with the values given above each line.}
\label{fig:esigmass}
\end{figure}

In Fig.~\ref{fig:esigmass} we plot $\left<\sige/\sigma_{e0}\right>$
versus $\Delta {\mathrm M}\BD/{\mathrm M}\BD_{\mathrm init}$, which
now shows a strong correlation.  For the correlation using \sige\BD\
we find a positive Spearman's rank correlation coefficient $r_s=0.91$
which is statistically significant at more than six sigma while using
\sige\B\ we find an even stronger correlation with $r_s=0.95$
corresponding to more than seven sigma.  This result is consistent
with the findings of \citet{Debattista2013} who showed that an
increase in disc mass within the bulge effective radius raises its
velocity dispersion.  The dotted lines in Fig.~\ref{fig:esigmass}
indicate different values of $(\sige/\sigma_{e0})^\beta$, where $\beta
= 4.24$ comes from the \Msige\ relation of \citet{Gueltekin2009b}.
These lines indicate the factor by which SMBHs must grow in order to
remain on the \Msige\ relation.  The factors get to be as large as
2-3.  A steeper \Msige\ relation (such as those of \citet{Graham2011}
and \citet{McConnell2013}) would require even larger growth factors.

\begin{figure*}
\centering
\begin{tabular}{cc}
\includegraphics[width=0.35\hsize,angle=270]{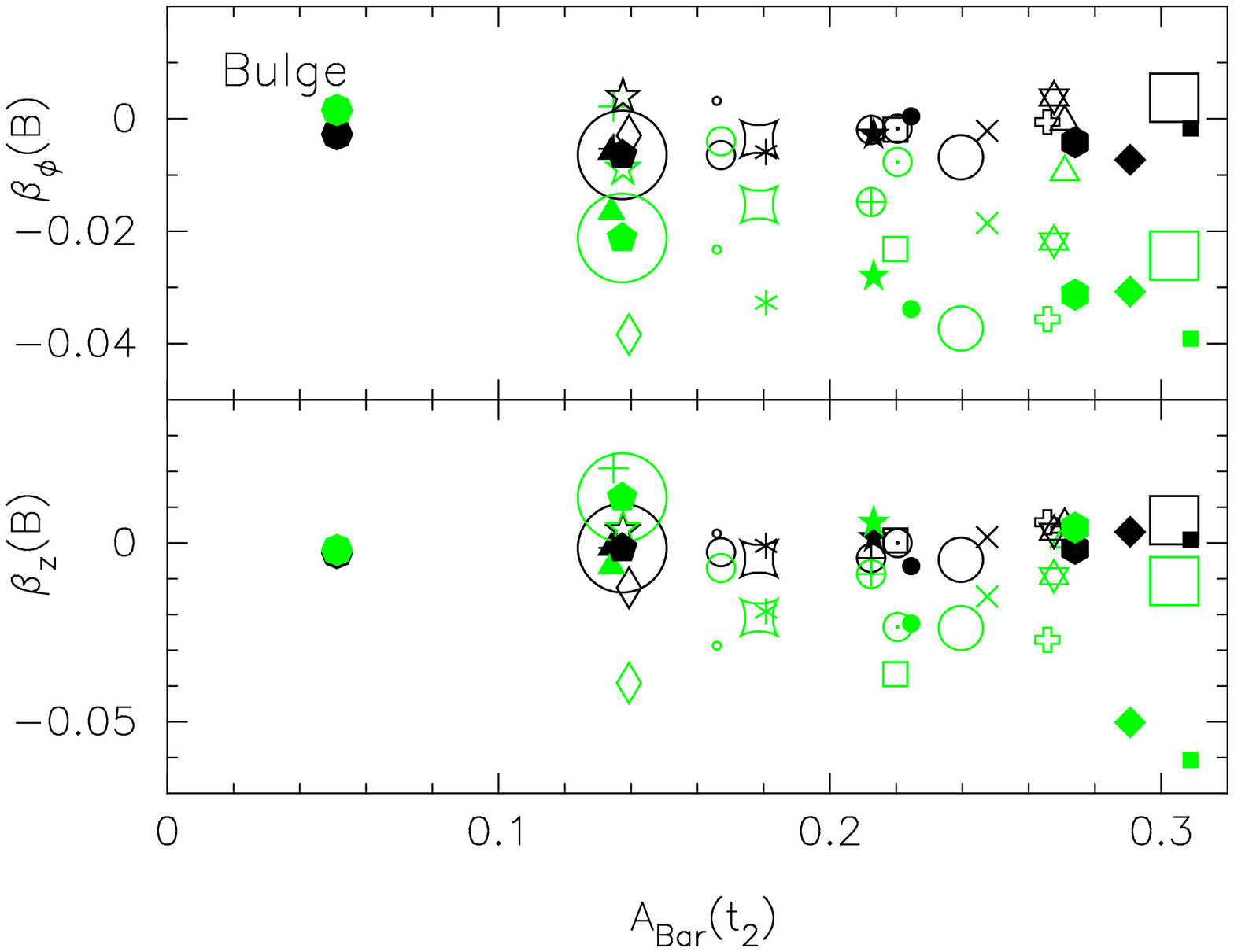} &
\includegraphics[width=0.35\hsize,angle=270]{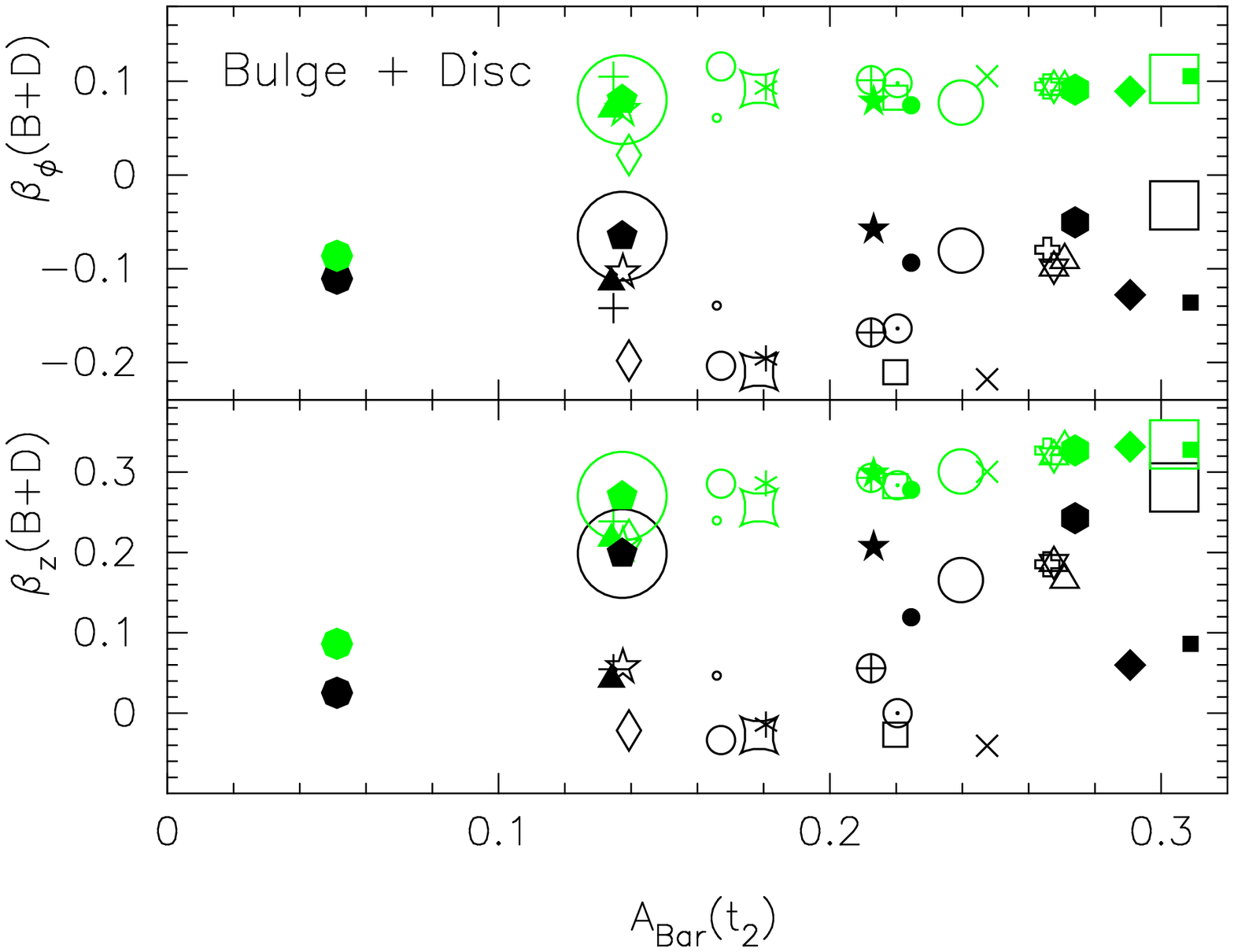} 
\end{tabular}
\caption{The anisotropies $\beta_{\phi}$ (top panels) and $\beta_z$
  (bottom panels) at $t_0$ (black) and $t_2$ (green) for bulge
  particles only (left panels) and for the bulge$+$disc particles
  (right panels) versus $\Abar\left(t_2\right)$. Note that the initial
  disc has no bar, so \Abar\ is zero; in order to show the evolution
  of the anisotropy we plot the initial anisotropies versus
  $\Abar\left(t_2\right)$.}
\label{fig:cylanisotropy}
\end{figure*}

The presence of a bar inherently leads to an anisotropic velocity
ellipsoid. We measure the velocity dispersions in cylindrical
coordinates $\sigma_u$, $\sigma_v$, $\sigma_w$ and obtain the
anisotropies $\beta_{\phi}=1-\sigma_v^2/\sigma_u^2$ and $\beta_{z} =
1-\sigma_w^2/\sigma_u^2$.  A positive value of $\beta_{\phi}$ or
$\beta_{z}$ implies that the radial velocity dispersion is larger than
the tangential or vertical one.  The initial bulge in all the models
is isotropic by construction \citep[classical bulges being well
described by flattened isotropic rotators][]{Kormendy1982,
  Davies1983}. Fig.~\ref{fig:cylanisotropy} shows that following the
formation of the bar, the velocity distributions of both the bulge and
the disc particles become anisotropic, with the degree of anisotropy
depending very weakly on the bar strength.  When only the bulge is
considered (left panel) all runs show only a slight tangential
anisotropy at $t_2$.  However when both the bulge and disc are
considered together, we measure a radial anisotropy up to
$\beta_\phi\BD\sim0.1$ and $\beta_z\BD$ reaching to $\sim0.35$.
Fig.~\ref{fig:cylanisotropy} also shows that $\beta_\phi\BD$ is
uncorrelated with \Abar, while $\beta_z\BD$ shows a very weak
correlation with \Abar. The lack of dependence of $\beta_\phi$ and
$\beta_z$ on bar strength is probably a result of the buckling
instability. As the degree of radial anisotropy increases the bar
becomes unstable to the buckling instability, which results in a
redistribution of kinetic energy and a decrease in anisotropy
\citep{Araki1987, Raha1991, Merritt1994}.

\begin{figure*}
\centering
\begin{tabular}{cc}
\includegraphics[width=0.37\hsize,angle=270]{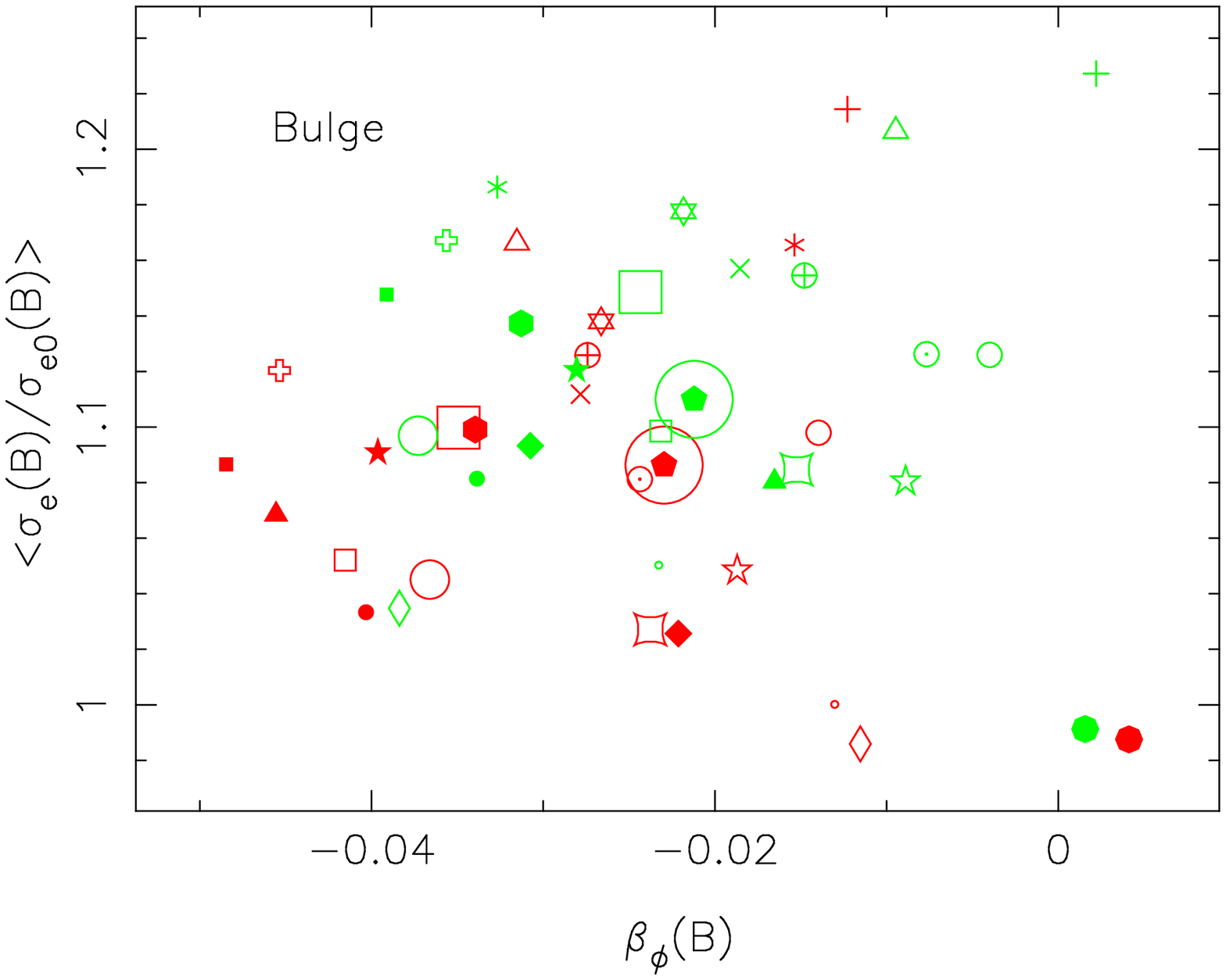} &
\includegraphics[width=0.37\hsize,angle=270]{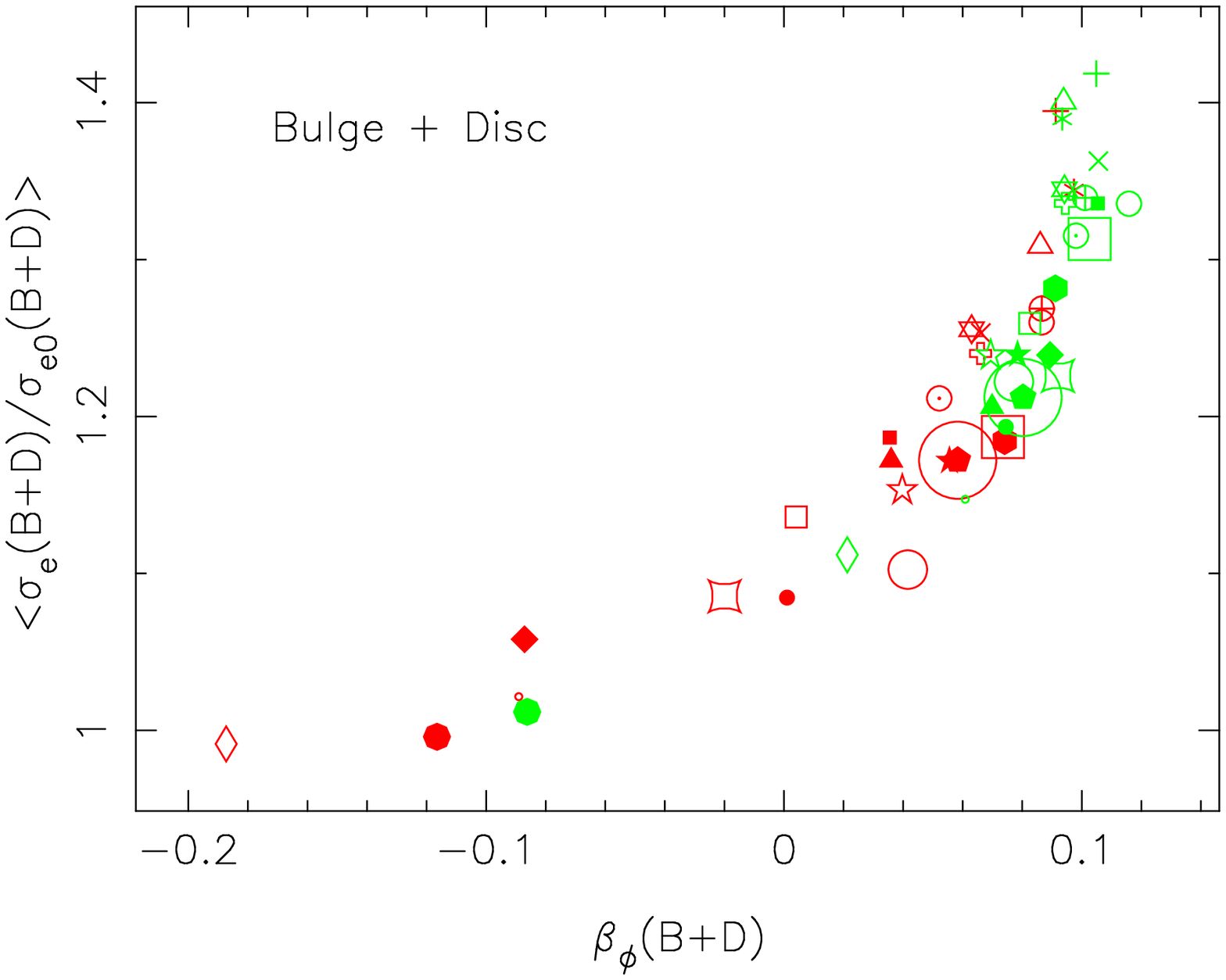} \\
\includegraphics[width=0.37\hsize,angle=270]{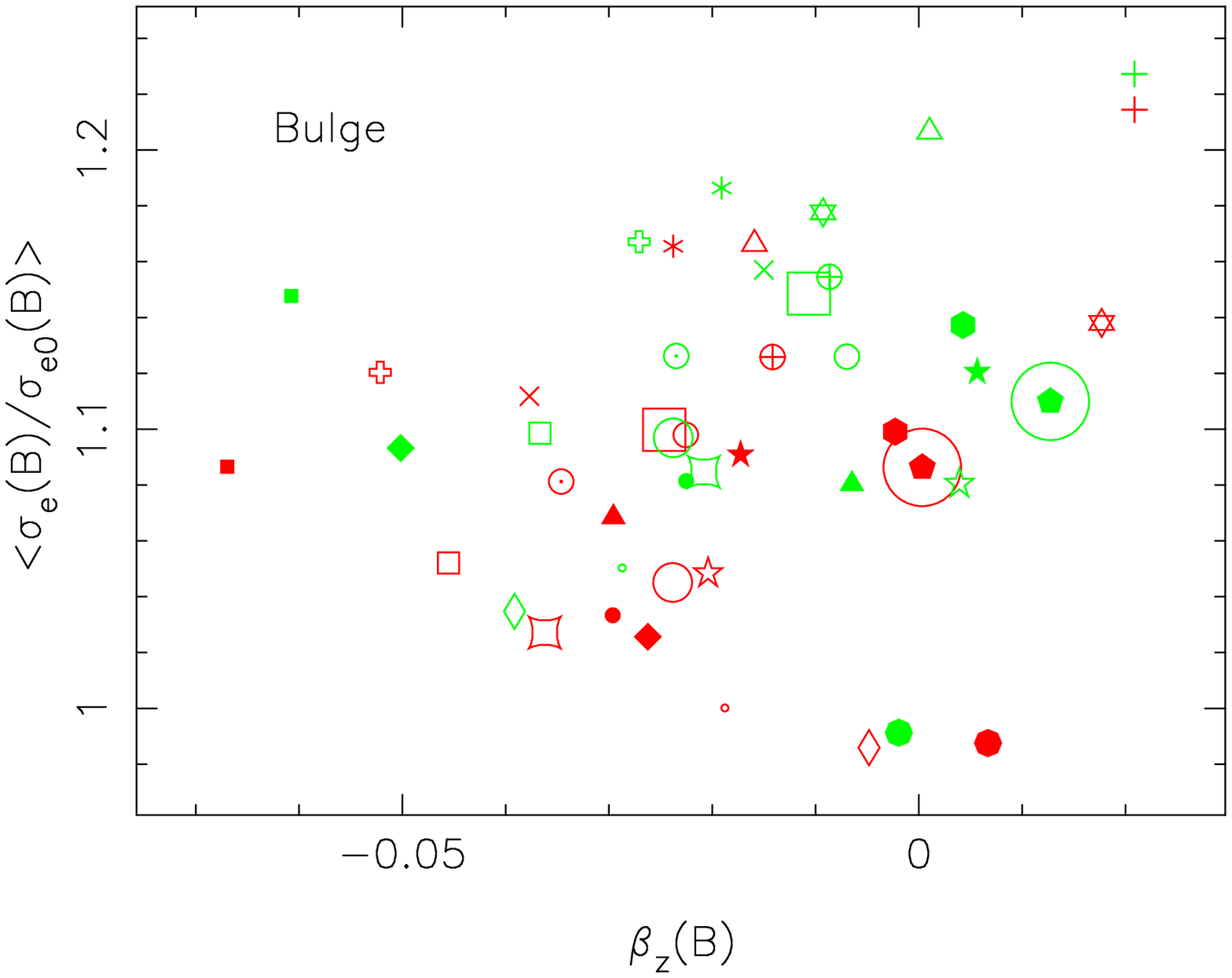} &
\includegraphics[width=0.37\hsize,angle=270]{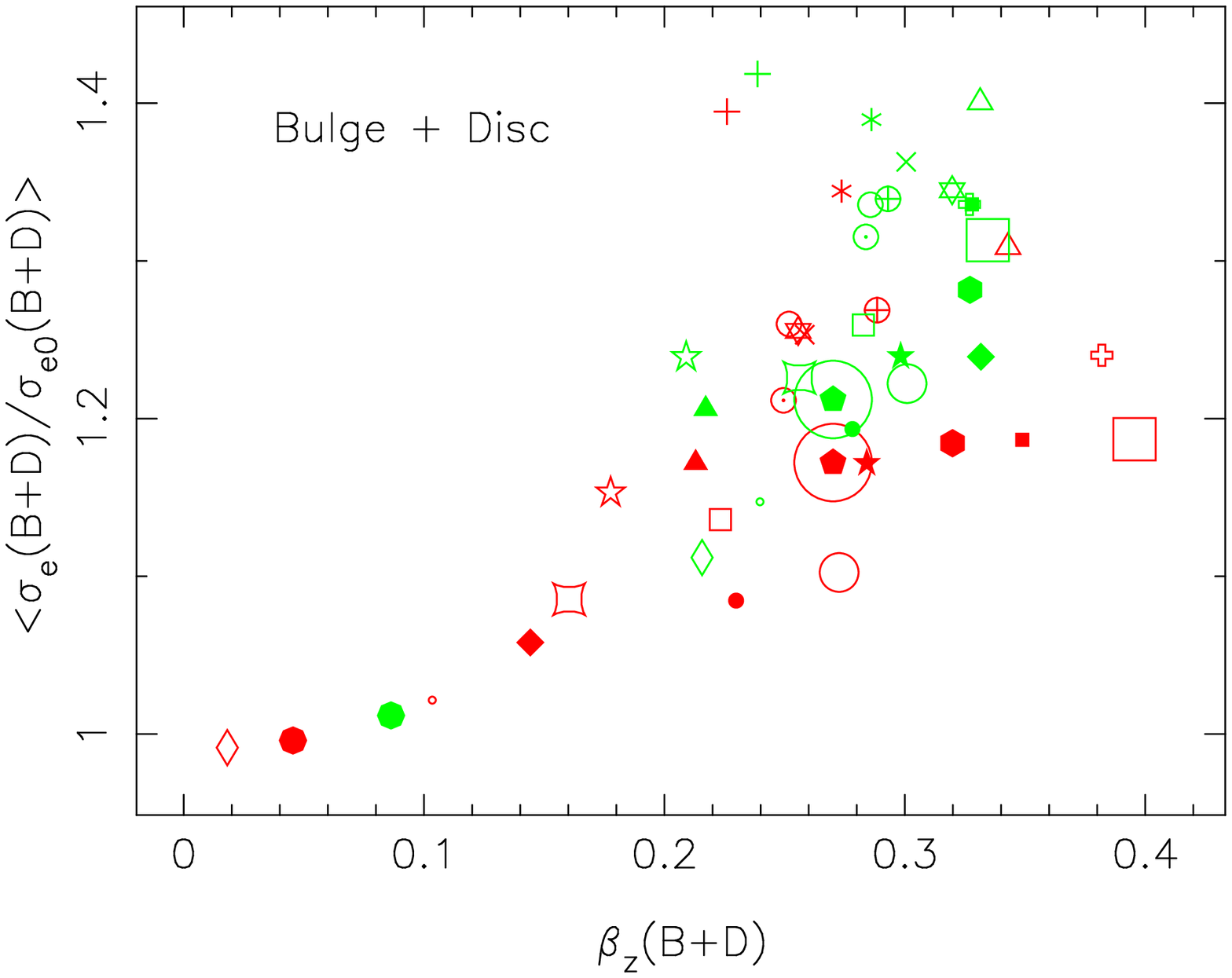} \\
\end{tabular}
\caption{The ratio of final to initial velocity dispersion,
  $\left<\sige/\sigma_{e0}\right>$, versus anisotropy $\beta_\phi$
  (top panels) and $\beta_z$ (bottom panels) for the bulge particles
  only (left panels) and for the bulge$+$disc particles (right
  panels). In all panels values at $t_1$ are red and at $t_2$ are
  green.}
\label{fig:anisotropy}
\end{figure*}

Fig.~\ref{fig:anisotropy} plots the orientation-averaged $\left<
  \sige/\sigma_{e0}\right>$ versus $\beta_\phi$ (top panels) and
$\beta_z$ (bottom panels).  No correlation is present for bulge
particles only (left panels).  However a very strong correlation is
present for bulge$+$disc particles and is stronger for $\beta_\phi\BD$
than for $\beta_z\BD$.  The increase in \sige\ is largest when
$\beta_\phi\BD$ is largest, implying that the orbits contributing to
the increased velocity dispersion are more radially biased.  Since the
correlation is absent when only bulge particles are considered the
disc particles must be primarily responsible for the increased
anisotropy \citep[e.g.][]{Saha2012}.  Fig.~\ref{fig:anisotropy} also
shows a temporal evolution, with the central regions becoming more
anisotropic and \sige\ increasing with time.

\begin{figure}
\includegraphics[width=0.8\hsize,angle=270]{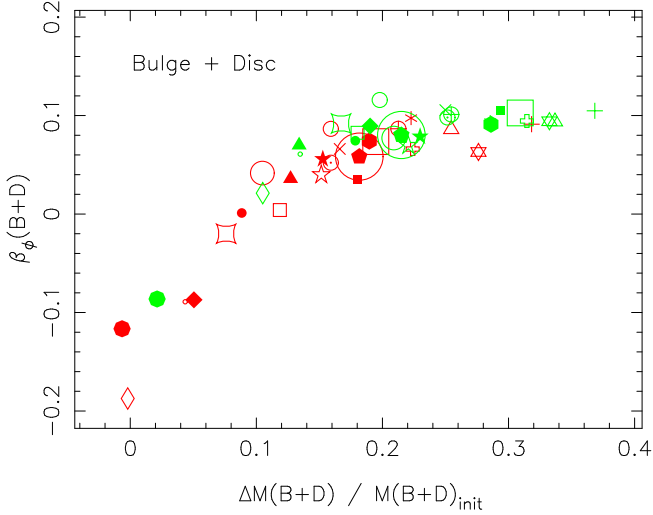} 
\caption{$\beta_\phi\BD$ versus the fractional changes in mass of the
  bulge$+$disc at $t_1$ (red) and $t_2$ (green).}
\label{fig:betamass}
\end{figure}

In Fig.~\ref{fig:betamass} we plot $\beta_\phi\BD$ versus the
fractional change in mass which shows a strong correlation but with
$\beta_\phi\BD$ saturating at $\sim 0.1$.  The Spearman coefficient
$r_s=0.67$; thus this correlation is weaker than the correlation
between $\left<\sige/\sigma_{e0}\right>$ and ${\rm \Delta M\BD/
  M\BD_{init}}$, which is presumably more fundamental.

\subsection{Effect of viewing orientation}

\begin{figure}
\includegraphics[width=0.8\hsize,angle=270]{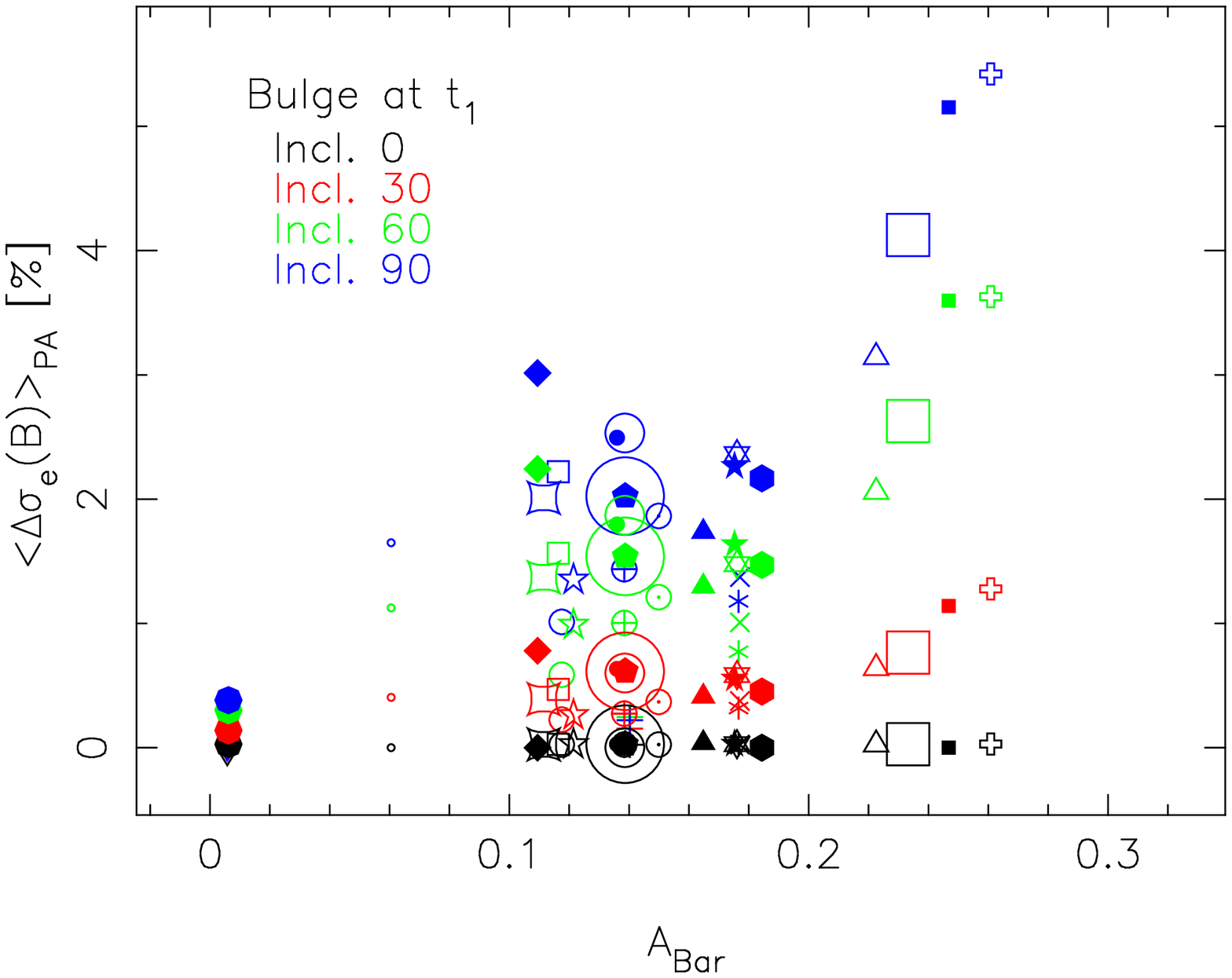}
\includegraphics[width=0.8\hsize,angle=270]{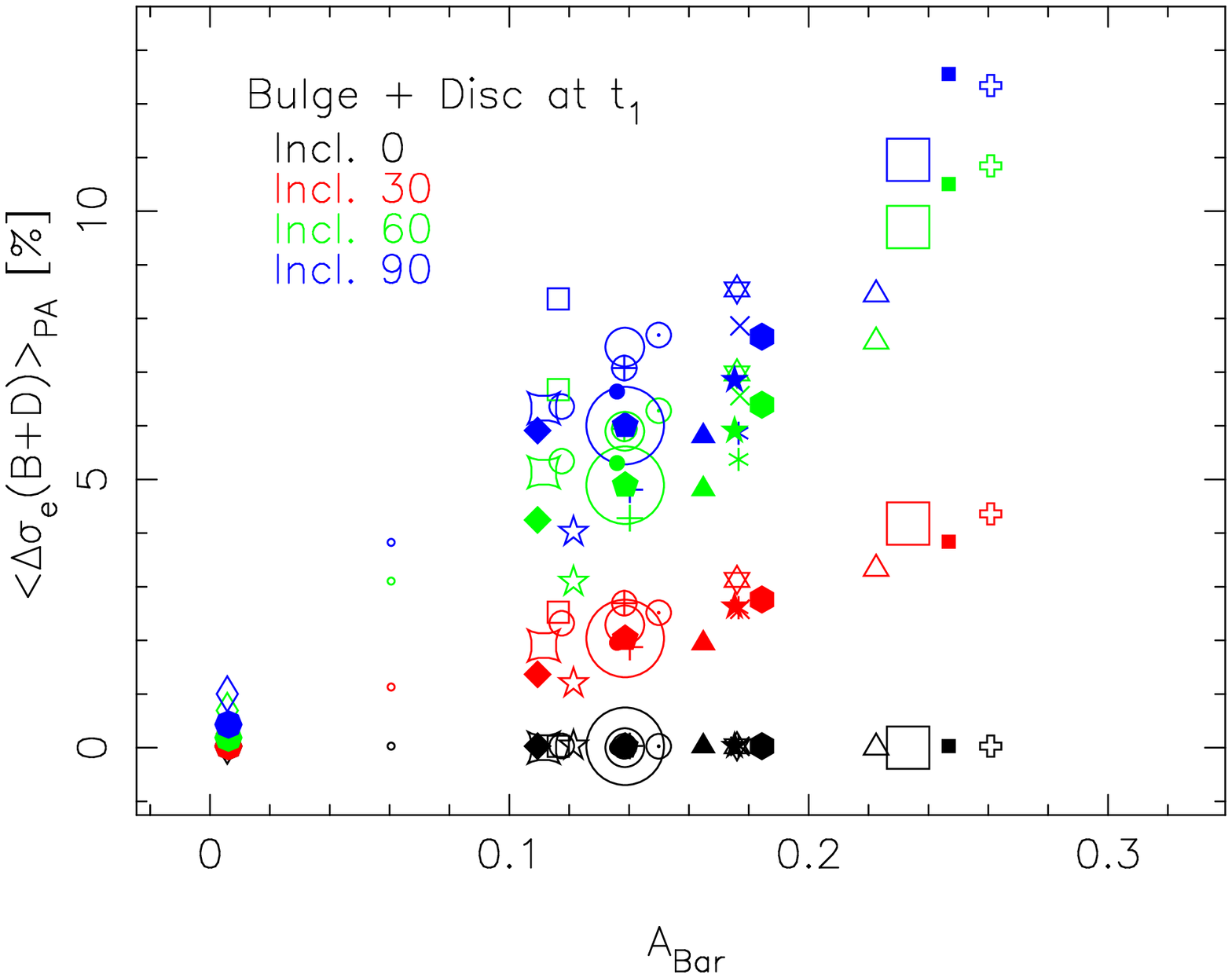}
\caption{The scatter in the velocity dispersion averaged over PA at
  fixed inclination, $\left< \Delta\sige \right>_{\rm PA}$, plotted
  versus bar amplitude at $t_1$.  $\left< \Delta\sige \right>$ is
  measured within \re\ for bulge (top panel) and bulge$+$disc
  particles (bottom panel).}
\label{fig:ScatterPA}
\end{figure}

Anisotropy increases the scatter in \sige, $\Delta \sige$
\citep{Graham2011}.  In Fig.~\ref{fig:ScatterPA} we show the scatter
in \sige\ by averaging it over position angles at fixed inclinations,
$\left< \Delta\sige\right>_{\rm PA}$.  We present results at $t_1$
which produces more fractional scatter than at $t_2$ in most cases
(the exception being in model 25 in which the bar is still very weak
at $t_1$).  For bulge particles $\Delta\sige$ is $\sim6\%$ but can be
as large as $\sim13\%$ for bulge$+$disc particles.  The scatter
increases with inclination and, at fixed inclination, with bar
strength.  In the face-on case, since we are measuring \sige\ within
circular apertures, $\Delta \sige=0$.

\section{Predicted Evolution of the \Msige\ relation}
\label{sec:msigma}

\begin{figure}
\includegraphics[width=0.8\hsize,angle=270]{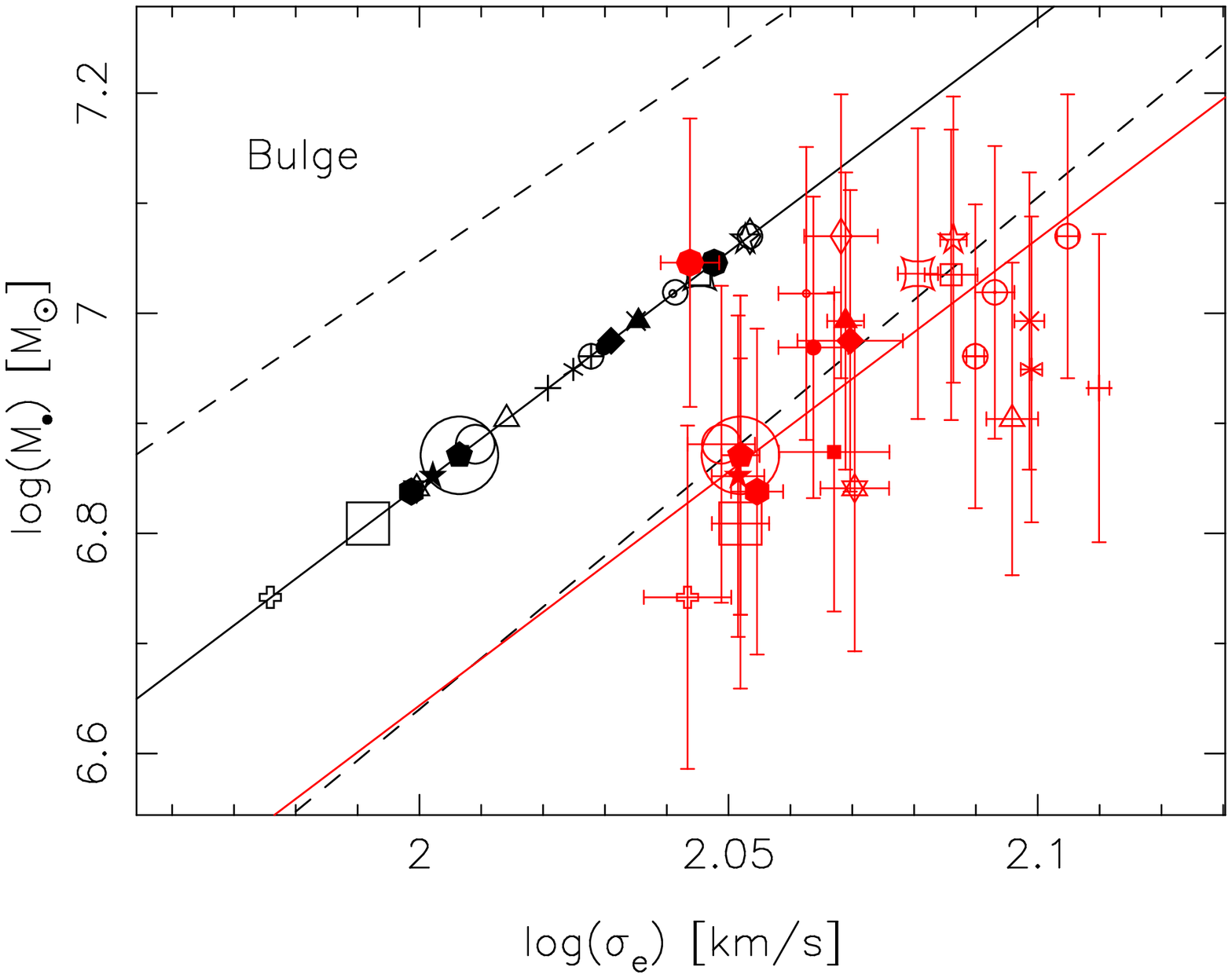}
\includegraphics[width=0.8\hsize,angle=270]{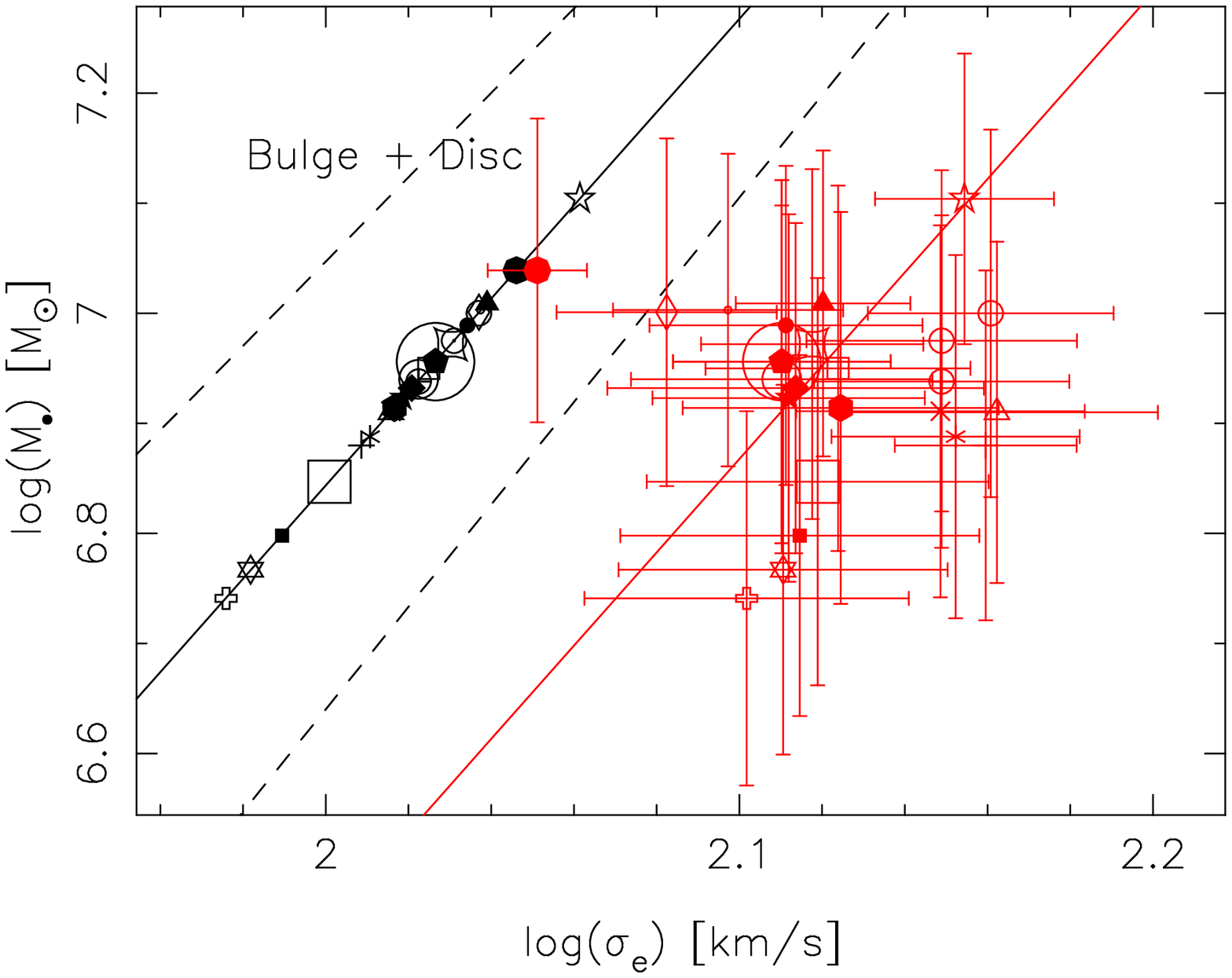} 
\caption{Using the \Msige\ relation of \citet{Gueltekin2009b} (solid
  black line, with dashed lines indicating the one $\sigma$
  uncertainty) we show the initial \sige\ and the corresponding \Mbh\
  (black symbols) for the simulations. Then assuming that \Mbh\ does
  not change, we plot \sige\ at $t_2$ (red symbols).  The red solid
  line shows a fit to the red points using MPFITEXY with slope fixed
  to $\beta=4.24$ to match the solid black line
  \citet{Gueltekin2009b}.  The top panel uses \sige\B\ while the
  bottom panel uses \sige\BD.  Note the different scale of the
  abscissa.  In both cases we find a substantial offset from the
  \Msige\ relation.}
\label{fig:msigrel}
\end{figure}

We have shown that the angular momentum redistribution of Fig.
\ref{fig:angmom} is a driver of major change in \sige.  Changes in
\sige\ can lead to displacements of a SMBH in the \Msige\ relation.
In this Section we estimate the effects of this \sige\ evolution on
the \Msige\ relation of barred galaxies.  Since the models we use do
not contain a SMBH we simply assume that \Mbh\ before the bar forms
satisfies the \Msige\ relation and explore what happens if \Mbh\ does
not change after the bar forms.

An increased \sige\ moves a SMBH to the right of the \Msige\ relation.
If the average fractional change in \sige\ is $\left<
  \sige/\sigma_{e0} \right>$, then we can write the \Msige\ relation,
assuming no \Mbh\ growth and that $\left<\sige/\sigma_{e0}\right>$ is
independent of $\sigma_{e0}$, as $\log{\Mbh}=\alpha + \beta
\log{\sige} - \beta \log{\left<\sige/\sigma_{e0}\right>}$.  Thus the
slope of the \Msige\ relation remains $\beta$, but the zero-point
changes by
\begin{eqnarray}
\delta\alpha=-\beta\log{\left<\sige/\sigma_{e0}\right>}
\label{eqn:offset}
\end{eqnarray}
\citep[see also][]{Debattista2013}.  Since $\left< \sige/\sigma_{e0}
\right> >1$, the resulting $\delta \alpha<0$, \ie\ the new \Msige\
relation will be offset below the \Msige\ relation of unbarred
galaxies.  We measure $\left<\sige/\sigma_{e0}\right> = 1.12 \pm 0.05$
for bulge particles only ($\left<\sige/\sigma_{e0}\right> = 1.27 \pm
0.12$ for bulge$+$disc particles).  This value of
$\left<\sige/\sigma_{e0}\right>$ would result in offsets in the range
$-\delta \alpha = 0.17$ to $0.27$ (bulge particles only) or $0.36$ to
$0.57$ (bulge$+$disc particles) for $\beta = 3.5-5.5$ if SMBHs do not
grow further.

In Fig.~\ref{fig:msigrel} we plot the models in the \Msige\ plane,
adopting $\beta=4.24$ from \citet{Gueltekin2009b}, at $t_0$ (before
the bars form) as black symbols and at $t_2$ (at the end of the
simulation) as red points.  We obtain \Mbh\ using \sige\B\ at $t_0$.
As expected, bar evolution without \Mbh\ growth shifts the models to
the right.  We measure the bar-induced offset by fitting the \Msige\
relation using MPFITEXY\footnote{http://purl.org/mike/mpfitexy}, which
implements the algorithm MPFIT \citep{Markwardt2009}, to obtain a
linear regression by minimising
\begin{eqnarray}
\chi^2=\sum_{i=1}^N\frac{\left(y_i-\alpha-\beta x_i\right)^2}{\epsilon_{x_i}^2 + \left( \epsilon_{y_i}^2+\eps^2\right)}
\end{eqnarray}
where $\eps$ is the intrinsic scatter, which is determined such that
the $\tilde{\chi}^2\lesssim1$ \citep{Tremaine2002}.  We fit the
\Msige\ relation for \sige\ at $t_2$ assuming that \Mbh\ remains
unchanged from $t_0$. For the errors in \sige\ we use $\Delta\sige$.
The errors in \Mbh\ are obtained from $\Delta\sige$ at $t_0$.  We
assume in these fits that the \Msige\ relation of barred galaxies has
the same $\beta = 4.24$ as do the unbarred galaxies, and therefore
hold $\beta$ fixed.  A significant offset develops regardless of
whether we measure $\sige\B$ or $\sige\BD$.  We find an offset
$\delta\alpha \simeq -0.20$ (see Table~\ref{tab:fits}).  Since the
scatter in the observed \Msige\ relation is generally estimated at
$\epsilon_0 = 0.3-0.45$ (see Table~\ref{tab:slopes}); an offset of
this magnitude is likely to be hard to measure.

We find a scatter of $\eps \simeq 0.1$ in the simulations due to the
different relative increases in central mass in the different models.
The full scatter predicted by the models includes that from viewing
orientation, which from the top-right panel of Fig.
\ref{fig:ScatterPA} we estimate at 0.05-0.09.  Thus the predicted
total increase in scatter relative to the intrinsic scatter in the
\Msige\ relation of unbarred galaxies is 0.11-0.13.

\begin{table}
\centering
\begin{minipage}{0.5\textwidth}
  \caption{Results of fitting the \Msige\ relation of the simulations
    assuming no \Mbh\ growth from a relation with $\beta=4.24$
    \citep{Gueltekin2009b} at $t_0$.  The offset is the difference
    between the zero-point of unbarred disc galaxies and from the fit
    to the barred models at $t_2$.}
\centering
\begin{tabular}{@{}cccc@{}}
\hline
\hline
Component    & $\alpha$  & Offset & Scatter   \\
             &           & [dex]  &           \\     
\hline
Bulge        & $7.92\pm0.03$ & -0.20 & 0.09 \\
Bulge$+$Disc & $7.91\pm0.03$ & -0.21 & 0.11 \\
\hline
\label{tab:fits}
\end{tabular}
\end{minipage}
\end{table}

\subsection{Residuals correlations}
\label{sec:fp}

\begin{figure}
\includegraphics[width=0.8\hsize,angle=270]{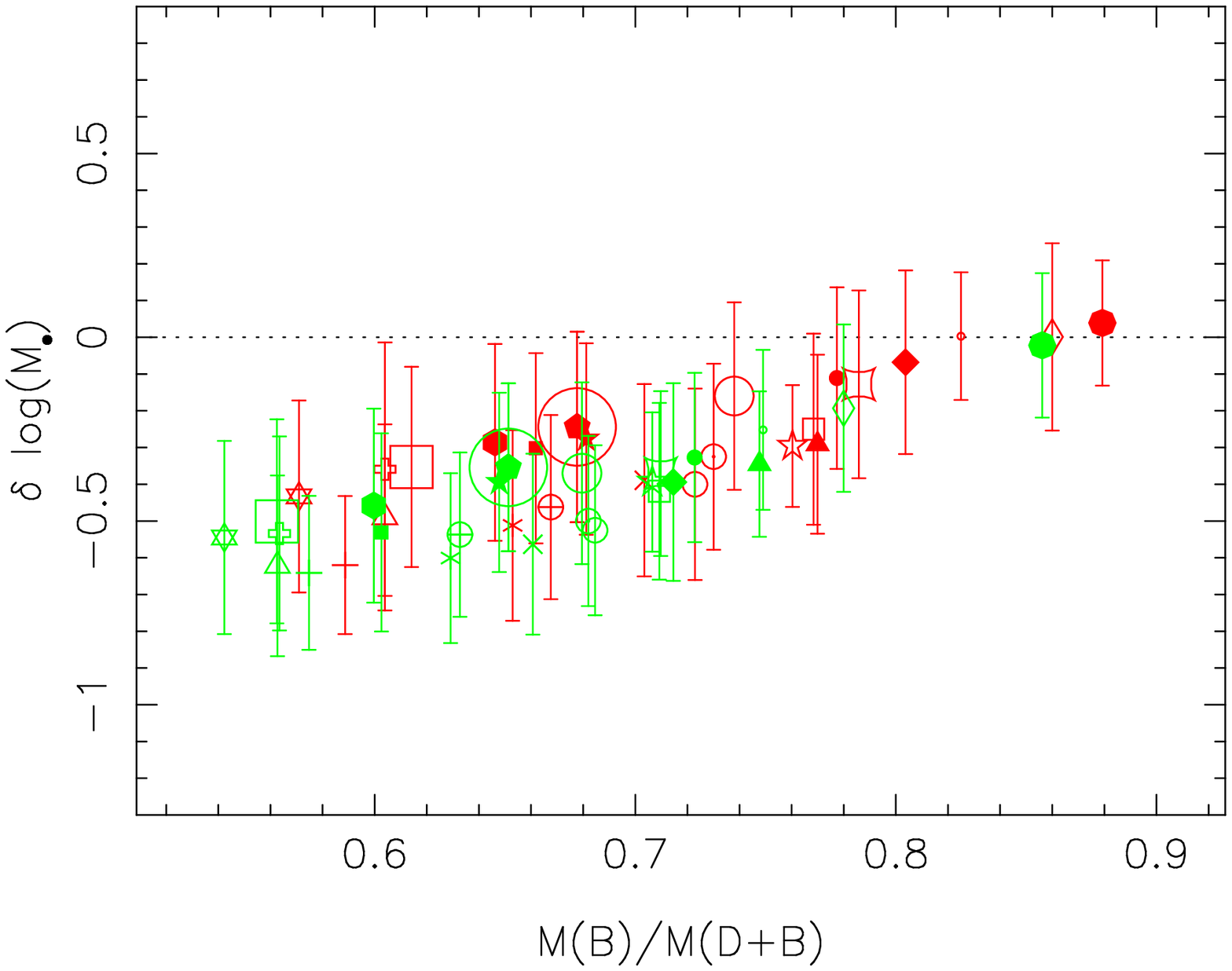} 
\includegraphics[width=0.8\hsize,angle=270]{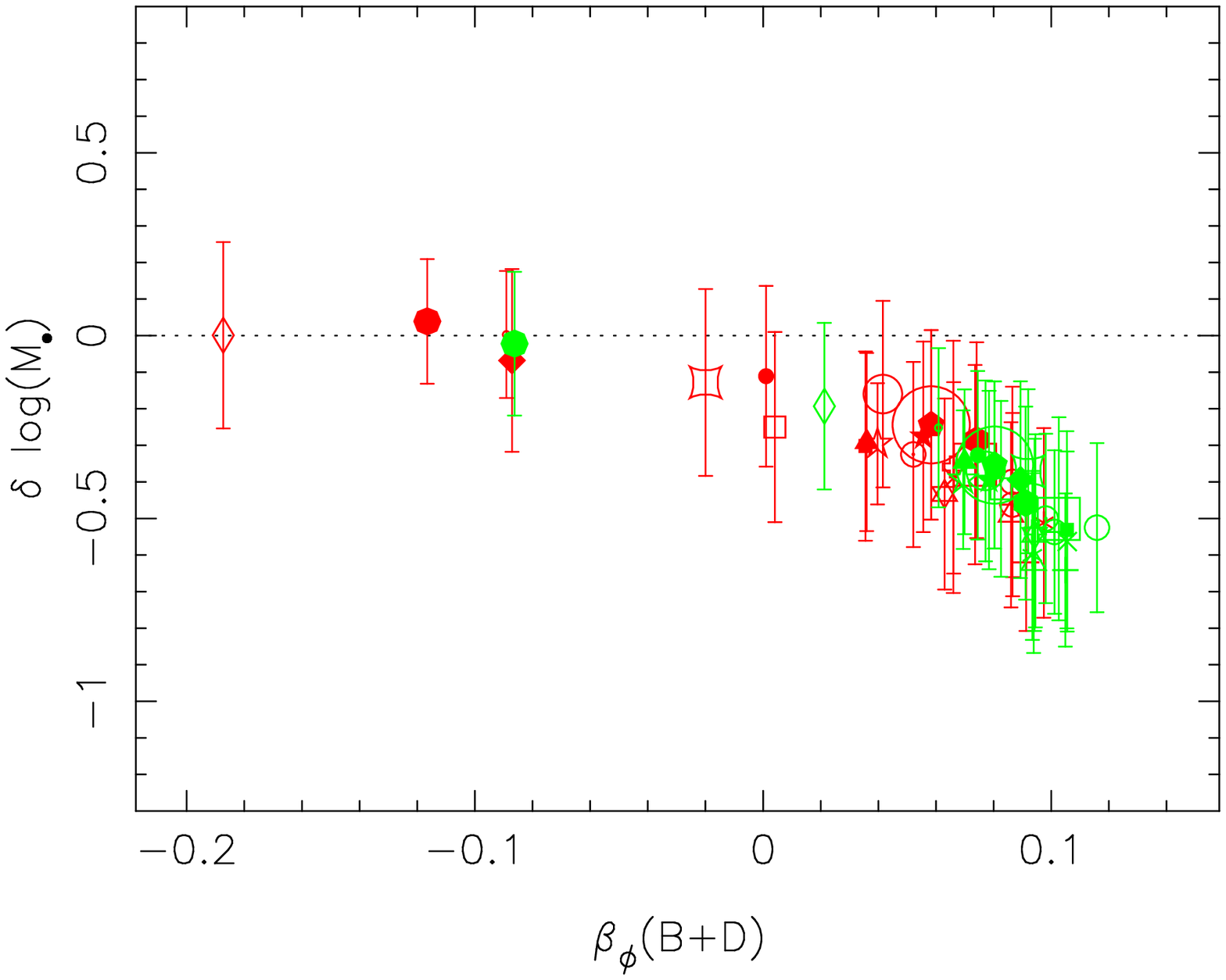}
\caption{The residuals of the simulations from the \Msige\ relation.
  Values are indicated in red at $t_1$ and in green at $t_2$.  Top:
  plotted as a function of $M\B/M\BD$.  Bottom: plotted as a function
  of $\beta_\phi\BD$.}
\label{fig:residuals}
\end{figure}

The main parameter that governs how much \sige\ increases, and thus
how far a barred galaxy strays from the \Msige\ relation, is the
fractional change of total mass within the bulge effective radius
\citep[][and Fig.~\ref{fig:esigmass} here]{Debattista2013}.
Unfortunately this is not directly observable because we can never
know what any galaxy looked like before the bar formed.  We have
searched for observationally accessible structural parameters that
correlate with $\Delta{\rm M\BD / M\BD_{init}}$.  It seems not
unreasonable to expect that ${\rm M\B / M\BD}$ within \re\ (note, this
is {\it not} the usual bulge-to-total ratio, $B/T$) correlates with
$\Delta{\rm M\BD / M\BD_{init}}$.  The top panel of Fig.
\ref{fig:residuals} therefore plots the \Msige\ residuals,
$\delta\log\Mbh$, as a function of ${\rm M\B / M\BD}$.  A clear
correlation is evident, with Spearman coefficient $r_s = 0.91$ at
$t_1$ and $r_s = 0.90$ at $t_2$.  Jointly, $t_1 + t_2$ produce a
correlation with $r_s = 0.89$, which is statistically significant at
more than six sigma.

Alternatively, we have already shown in Figs. \ref{fig:anisotropy}
and \ref{fig:betamass} that the anisotropies correlate with the change
in \sige\ and with the fractional mass change.  Thus rather than a
structural parameter, a kinematic one may provide an alternative
indication of the offset of a barred galaxy from the \Msige\ relation.
Note however that Fig.~\ref{fig:cylanisotropy} shows that the
anisotropy of the bulge component only is not much changed by the bar,
and Fig.~\ref{fig:anisotropy} shows that $\beta_\phi\B$ and
$\beta_z\B$ do not correlate with $\left<\sige/\sigma_{e0}\right>$.
It is therefore the anisotropy of the bulge$+$disc that must be
measured to determine the offset.  The bottom panel of Fig.
\ref{fig:residuals} plots $\delta \log \Mbh$ as a function of
$\beta_\phi\BD$; we find a Spearman coefficient $r_s = -0.89$ at $t_1$
and $r_s = -0.82$ at $t_2$, with a joint ($t_1+t_2$) $r_s = -0.85$
corresponding to almost six sigma significance.  The strength of the
correlation between $\delta\log\Mbh$ and $\beta_z\BD$, instead, has
$r_s = -0.66$ at $t_1$ and $r_s = -0.61$ at $t_2$, with a joint $r_s =
-0.64$ corresponding to more than four sigma significance.

A concern with the correlation between ${\rm M\B / M\BD}$ and
$\Delta{\rm M\BD / M\BD_{init}}$ is that it could be weakened if
models with a larger range of $B/D$ at $t_0$ were included.  Moreover,
while we can easily compute ${\rm M\B / M\BD}$ in the simulations, the
same quantities may be non-trivial in observations because once the
bar forms the density profile of the disc need no longer be an
exponential extending to small radii \citep[e.g.][]{Debattista2006}.
The main limitation of using $\beta_\phi\BD$ to measure the residuals
instead is that it tends to saturate, at least in these collisionless
simulations.  In addition, in a companion paper, \citet{Brown2013}
show that the growth of a central massive object inside a barred
galaxy will tend to isotropise the velocity distribution.
Nonetheless, we propose that modelling the velocity anisotropy is
worthwhile in order to understand the offsets of barred galaxies from
the \Msige\ relation.

\section{Comparison with observations}
\label{sec:obs}

Evidence for an offset in the \Msige\ relation of barred galaxies has
been presented by a number of authors \citep{Hu2008, Graham2008b,
  Graham2011}. Guided by the results above, here we retest for an
offset by fixing the slope of the barred \Msige\ relation to that of
unbarred galaxies and measuring the zero-point.  

\subsection{Sample selection}

We have assumed that SMBHs are present in bulges and satisfy the
\Msige\ relation before the bar forms, and that the bulges do not grow
any further once the bar forms.  Both these assumptions imply that
classical bulges are more suited to compare with the simulations.
Classical bulges share structural and kinematical properties with
elliptical galaxies \citep[\eg][]{Wyse1997, Kormendy2004,
  Gadotti2009b}, with both types of spheroids appearing in similar
positions on the fundamental plane defined by the central velocity
dispersion, the central surface brightness, and the effective radius
\citep{Bender1992}.  Pseudo bulges instead are more closely related to
the discs of their host galaxy \citep[see][for a
review]{Kormendy2004}.  While elliptical galaxies and classical bulges
are believed to form via mergers of galaxies and accretion
\citep{Eggen1962, Tremaine1975, Searle1978, Kauffmann1993, Baugh1996,
  vandenBosch1998, Naab2007}, pseudo bulges are thought to form via
secular processes in the disc which are driven by non-axisymmetric
structures such as bars and spirals \citep{Combes1981, Combes1990,
  Raha1991, Norman1996, Courteau1996, Bureau1999, Debattista2004,
  Athanassoula2005, Drory2007}.  The difference between classical and
pseudo bulges is reflected also in their SMBH scaling relations.
\citet{Hu2008} and \citet{Debattista2013} found that SMBHs in
elliptical galaxies and in classical bulges follow a similar \Msige\
relation.  Pseudo bulges instead either have a significant offset from
this \Msige\ relation \citep{Hu2008} or no \Msige\ relation at all
\citep{Kormendy2011a}.

In this work we therefore distinguish observed galaxies by whether
they contain a classical or a pseudo bulge.  We use data from the
literature to compile samples of unbarred classical bulges and barred
classical bulges with \Mbh\ measurements.  Purely for the sake of
comparison we also compile a sample of barred pseudo bulges.  The
final sample of galaxies is listed in Table~\ref{tab:bhmassamples}.
Our sample of \Mbh\ and \sige\ measurements is primarily drawn from
the compilation of \citet{McConnell2013}, with one galaxy (NGC~7457)
from \citet{Gueltekin2009b} and another (NGC~3414) from
\citet{Graham2013}.  We largely rely on the {\it morphological}
classification of \citet{Fisher2008, Fisher2010, Fisher2011}.  For
some bulges our classification is based solely on S\'{e}rsic index
$n>2$ of a bulge$+$disc decomposition.  For these cases we use
unpublished fits provided to us by David Fisher supplemented by fits
by \citet{Beletsky2011}, \citet{Rusli2011}, \citet{Fabricius2012} and
\citet{Krajnovic2013}.  We classify the Milky Way as having a pseudo
bulge although this is controversial; the bulge$+$disc decomposition
is based on the model of \citet{Bissantz2002}.  From the sample of
disc galaxies in \citet{McConnell2013}, we exclude those where the
bulge classification is unknown or where the galaxy is unbarred and
hosts a pseudo bulge.  We exclude NGC~4826 from our sample because of
confusion over its bulge type \citep{Fabricius2012}, and NGC~2549
because the only available profile fit uses only a single S\'ersic
\citep{Krajnovic2013}.  We exclude the remaining ten barred galaxies
because no bulge$+$disc fits are available but include them in a
separate unclassified bulge barred galaxy sample.
Table~\ref{tab:bhmassamples} presents our samples of galaxies,
consisting of twelve unbarred galaxies with classical bulges, five
barred galaxies with classical bulges and nine barred galaxies with
pseudo bulges.

\begin{table*}
  \begin{center}
    \begin{tabular}{lccllll} 
      \hline
      \multicolumn{1}{l}{Galaxy} &
      \multicolumn{1}{c}{Type} &
      \multicolumn{1}{c}{Bulge} &
      \multicolumn{1}{c}{S\'ersic} &
      \multicolumn{1}{c}{B/D} &
      \multicolumn{1}{c}{$M_\bullet$} &
      \multicolumn{1}{c}{$\sigma_e$} \\
      \multicolumn{1}{l}{} &
      \multicolumn{1}{c}{} &
      \multicolumn{1}{c}{classification$^a$} &
      \multicolumn{1}{c}{Index n} &
       \multicolumn{1}{c}{} &
      \multicolumn{1}{c}{$10^8 M_\odot$} &
      \multicolumn{1}{c}{km s$^{-1}$} \\[-0.5ex]
      \hline
Unbarred \\
      \hline
NGC1332$^{1,7,13}$ & S0 & C & $2.36$ & $0.79$ & $15\pm2$ & $328\pm16$ \\
M81 (NGC3031)$^{1,4b}$ & Sb & C & $3.88\pm0.23$ & $0.59$ & $0.8^{+0.2}_{-0.11}$ & $143\pm7$ \\
NGC3115$^{1,4a}$ & S0 & C & $3.89\pm0.32$ & $1.63$ & $8.9^{+5.1}_{-2.7}$ & $230\pm11$ \\
NGC3245$^{1,4a}$ & S0 & C & $3.82\pm0.34$ & $1.44$ & $2.1^{+0.5}_{-0.6}$ & $205\pm10$ \\
NGC3414$^{2,6,10,13}$ & S0 & C & $2.3\pm0.9$  & $0.52$ & $2.4\pm0.3$ & $ 236.8\pm7.5$ \\
NGC3585$^{1,4d,13}$ & S0 & C & $3.49$ & $2.23$ & $3.3^{+1.5}_{-0.6}$ & $213\pm10$ \\
NGC3998$^{1,4d,8,13}$ & S0 & C & $4.1$ & $1.38$ & $8.5\pm0.7$ & $272\pm14$ \\
NGC4026$^{1,4d,8,13}$ & S0 & C & $2.46$ & $0.47$ & $1.8^{+0.6}_{-0.3}$ & $180\pm9$ \\
NGC4342$^{1,4d,8,13}$ & S0 & C & $4.84$ & $1.63$ & $4.6^{+2.6}_{-1.5}$ & $225\pm11$ \\
NGC4564$^{1,4a}$ & S0 & C & $3.70\pm0.66$ & $1.5$ & $0.88\pm0.24$ & $162\pm8$ \\
NGC4594$^{1,4c}$ & Sa & C & $6.2\pm0.6$ & $1.04$ & $6.7^{+0.5}_{-0.4}$ & $230\pm12$ \\
NGC7457$^{3,4b}$ & S0 & C & $2.72\pm0.4$ & $0.15$ & $0.041^{+0.012}_{-0.017}$ & $67\pm3$ \\
      \hline
Barred \\
      \hline
M31 (NGC224)$^{1,4c,12}$ & SBb & C & $2.1\pm0.5$ & $0.92$ & $1.4^{+0.8}_{-0.3}$ & $160\pm8$ \\
NGC1023$^{1,4b}$ & SB0 & C & $2.47\pm0.34$ & $0.54$ & $0.4\pm0.04$ & $205\pm10$ \\
NGC1316$^{1,5,13}$ & SB0 & C & $2.9$ & $0.59$ & $1.7\pm0.3$ & $226\pm11$ \\
NGC4258$^{1,4c,11}$ & SABbc & C & $2.80\pm0.28$ & $0.12$ & $0.367\pm0.001$ & $115\pm10$ \\
NGC4596$^{1,4d,13}$ & SB0 & C & $3.61$ & $1.04$ & $0.84^{+0.36}_{-0.25}$ & $136\pm6$ \\
MW$^1$ & SBbc & P & $1.0 $& $0.12$ & $0.041\pm0.006$ & $103\pm20$ \\
NGC1068$^{2,15}$ & SBb & P & - & - & $0.084\pm0.003$ & $151\pm7$ \\
NGC1300$^{1,4a}$ & SB(rs)bc & P & $1.61\pm 0.39$ & $0.09$ & $0.71^{+0.34}_{-0.18}$ & $218\pm10$ \\
NGC2787$^{1,4c,15}$ & SB0 & P & $2.6\pm0.5$ & $1.38$ & $0.41^{+0.04}_{-0.05}$ & $189\pm9$ \\
NGC3227$^{1,4d,15}$ & SBa & P & $2.49$ & $0.18$ & $0.15^{+0.05}_{-0.08}$ & $133\pm12$ \\
NGC3368$^{1,4b,9}$ & SBab & P & $1.63\pm0.18$ & $0.35$ & $0.076^{+0.016}_{-0.015}$ & $122^{+28}_{-24}$ \\
NGC3384$^{1,4b}$ & SB0 & P & $1.42\pm0.2$ & $0.49$ & $0.11^{+0.05}_{-0.05}$ & $143\pm7$ \\
NGC3489$^{1,4b}$ & SAB0 & P & $1.47\pm0.28$ & $1.45$ & $0.06^{+0.008}_{-0.009}$ & $100^{+15}_{-11}$ \\
NGC7582$^{1,4d,13}$ & SBab & P & $0.91$ & $0.10$ & $0.55^{+0.16}_{-0.11}$ & $156\pm19$ \\
      \hline
Barred galaxies with no bulge classification\\
      \hline
IC2560$^{2,14}$ & SBb & - & - & - & $0.044^{+0.044}_{-0.022}$ & $144$ \\
NGC253$^{2,10}$ & SBc & - & - & - & $0.10^{+0.10}_{-0.05}$ & $109\pm20$ \\
NGC2273$^{1}$ & SBa & - & - & - & $0.078\pm0.004$ & $144^{+18}_{-15}$ \\
NGC2549$^{1}$ & SB0 & - & - & - & $0.14^{+0.01}_{-0.04}$ & $145\pm7$ \\
NGC2778$^{2,10}$ & SB0 & - & - & - & $0.15^{+0.09}_{-0.1}$ & $161.7\pm3.2$ \\
NGC3393$^{1}$ & SBa & - & - & - & $0.33\pm0.02$ & $148\pm10$ \\
NGC4151$^{2,15}$ & SBab & - & - & - & $0.65\pm0.07$ & $119\pm26$ \\
NGC4945$^{2,10}$ & SBcd & - & - & - & $0.014^{+0.014}_{-0.007}$ & $127.9\pm19.1$ \\
NGC6323$^{1}$ & SBab & - & - & - & $0.098\pm0.001$ & $158^{+28}_{-23}$ \\
UGC3789$^{1}$ & SBab & - & - & - & $0.108^{+0.006}_{-0.005}$ & $107^{+13}_{-12}$ \\
      \hline
    \end{tabular}
  \end{center}
  \caption[]{
Published values for the black hole mass and bulge velocity dispersion 
for the galaxies plotted in Fig.~\ref{fig:obsmsig}. \\
$^a$ C= Classical, P=Pseudo. Classification take into account S\'ersic index as well as bulge morphology except where indicated. \\
$^1$ Black hole masses and velocity dispersion data from \citet{McConnell2013}. \\
$^2$ Black hole masses and velocity dispersion data from \citet{Graham2013}. \\
$^3$ Black hole masses and velocity dispersion data from \citet{Gueltekin2009b}. \\
$^{4a}$ Bulge/disc decompositions from \citet{Fisher2008}.\\
$^{4b}$ Bulge/disc decompositions from \citet{Fisher2010}.\\
$^{4c}$ Bulge/disc decompositions from \citet{Fisher2011}.\\
$^{4d}$ Bulge/disc decompositions from David Fisher (private communication).\\
$^5$ Bulge/disc decompositions from \citet{Beletsky2011}.  \\
$^6$ Bulge/disc decompositions from \citet{Krajnovic2013}. \\
$^7$ Bulge/disc decompositions from \citet{Rusli2011}. \\
$^8$ S\'ersic indices found by \citet{Krajnovic2013} are $<2$. Those from \citet{Fisher2010} are preferred because their data is based on HST observations which are of higher resolution than those of \citet{Krajnovic2013} which are based on SDSS data and from imaging with the Wide Field Camera (WFC)
mounted on the 2.5-m Isaac Newton Telescope. Bulge/disc decompositions based on lower resolution observations tend to give lower values of n therefore the value of n given for NGC3414 may be trusted to classify its bulge as any error will tend to lower the value of n but it is still $>2$. \\
$^{9}$ S\'ersic index found by \citet{Fabricius2012} is $2.46\pm0.77$ but morphologically classified as pseudo bulge. \\
$^{10}$ Source data from HyperLeda \citep{Paturel2003}. \\
$^{11}$ Described as pseudo bulge (possibly classical) in \citet{Fisher2010} but we use the updated classification of \citet{Fisher2011}. \\
$^{12}$ For barred classification see \citet{Athanassoula2006}. \\
$^{13}$ Classified only on basis of S\'ersic index. $n>2$ implies a classical bulge, $n<2$ implies a pseudo bulge. \\
$^{14}$ Velocity dispersion from \citet{CidFernandes2004}; for fitting purposes we assume an uncertainty of $\pm20\kms$. \\
$^{15}$ Bulge classification (and \sige\ for NGC 1068) from \citet{Kormendy2011a}.
\label{tab:bhmassamples} }
\end{table*}

\subsection{The \Msige\ relation of unbarred classical bulges}

\begin{figure}
\centering
\includegraphics[width=\hsize]{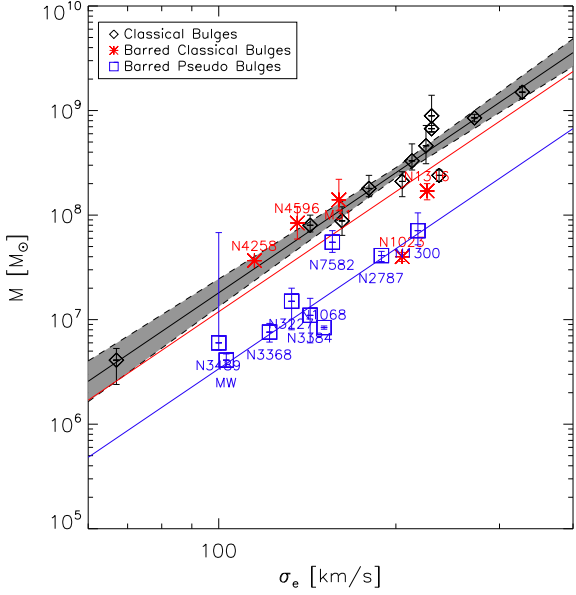}
\caption{The \Msige\ relations of classical bulges in unbarred
  galaxies, and of classical and pseudo bulges in barred galaxies from
  Table~\ref{tab:bhmassamples}. The solid black line shows the linear
  regression of the unbarred classical bulges, while the shaded region
  bounded by the dashed black lines shows the one $\sigma$
  uncertainty.  The solid red and blue lines show fits for the
  classical and pseudo bulges in barred galaxies with slope fixed to
  that for unbarred classical bulges.}
\label{fig:obsmsig}
\end{figure}

\begin{table}
\centering
\begin{minipage}{0.5\textwidth}
  \caption{Fit results: Using the data of Table~\ref{tab:bhmassamples}
    to fit the \Msige\ relation of only classical bulges in unbarred
    galaxies. We then fix the resulting slope and fit only the
    zero-point to obtain the offset of classical bulges and pseudo
    bulges in barred galaxies. N is the number of galaxies in each
    sample.}
  \centering
\begin{tabular}{@{}lcccc@{}}
\hline
\hline
Sample                    & N & $\beta$ & $\alpha$ & $\epsilon_0$ \\
                          &   &         &          &   [dex]    \\ \hline
Unbarred classical bulges & 12& $3.82\pm0.28$ & $8.41\pm0.05$ & $0.14$ \\ \hline
Barred classical bulges   & 5 & $3.82$        & $8.22\pm0.19$ & $0.42$ \\
Barred pseudo bulges      & 9 & $3.82$        & $7.68\pm0.07$ & $0.14$ \\
Barred classical bulges+  &   &               &               &        \\
\hspace{1em} unclassifieds& 15& $3.82$        & $7.88\pm0.14$ & $0.49$ \\
\hline
\label{tab:obsfits}
\end{tabular}
\end{minipage}
\end{table}

Using MPFITEXY, we first fit the \Msige\ relation for the unbarred
classical bulges.  The full parameters of the fit are listed in
Table~\ref{tab:obsfits}; we obtain a slope $\beta = 3.82 \pm 0.28$.
Remarkably this sample of twelve galaxies chosen purely by their
morphology have a quite small scatter of only 0.14 dex.  While this
can merely be due to small number statistics, the wide range of \sige\
considered, $67 \leq \sige \leq 328$~\kms, hints that our approach of
selecting sub-samples based on bulge type is reasonable.

The fit we obtain for unbarred classical bulges is in fairly good
agreement with the fit to elliptical galaxies (excluding brightest
cluster galaxies) of \citet{Debattista2013}: $(\alpha,\beta) =
(8.21\pm 0.07,4.06\pm0.40)$.  If we fix $\beta = 4.06$ and refit these
12 unbarred galaxies with classical bulges, we obtain $\alpha = 8.41
\pm 0.05$, which confirms the lack of an offset between ellipticals
and unbarred classical bulges.  \citet{Debattista2013} used this
result to argue for the need of SMBHs in classical bulges to grow
along with the discs.

\subsection{The \Msige\ relations of barred galaxies}

We then fix $\beta=3.82$ and fit the \Msige\ relation of barred
galaxies with classical bulges.  We obtain $(\alpha,~\epsilon_0) =
(8.22\pm0.19, 0.42)$, or $\delta \alpha=-0.19 \pm 0.20$~dex.  This
offset is smaller than the $\delta \alpha=-0.5$~dex found by
\citet{Graham2011} (who however considered all barred galaxies, not
just those with classical bulges).  The offset we find is consistent
with the one predicted by the models (see Table~\ref{tab:fits}).
However with this small sample it is also consistent with no offset.
The offset is largely driven by NGC~1023, while NGC~1316 also falls
below the unbarred \Msige\ relation.  The scatter is more than twice
as large as predicted by the models $\sqrt{0.14^2 + 0.11^2} = 0.18$
dex.  The main reason for this discrepancy is very likely the narrow
range of models we have considered, which reduces the predicted
scatter.  Moreover, if some SMBHs in barred galaxies are able to grow
again, returning to the fiducial \Msige\ relation, then this would
further increase the scatter (relative to the offset relation).
Indeed M31, NGC~4258 and NGC~4596 are all close to the \Msige\
relation of unbarred classical bulges.  Of these, M31 and NGC~4258
both have gas.  NGC~4258 is also the only galaxy in this sample with a
weak bar.  Finally, the observed scatter for barred galaxies may be
enhanced by modelling uncertainties.  For instance, in the presence of
a bar, the growth of a SMBH results in a larger increase in \sige\
than it would in an axisymmetric galaxy.  However the kinematics of
stars associated with the bar results in a smaller and frequently
negative Gauss-Hermite coefficient $h_4$.  \citet{Brown2013} argue
that using an axisymmetric stellar dynamical modeling to measure SMBH
masses in barred galaxies could result in an overestimate of the
derived \Mbh\ since low/negative $h_4$ values primarily result from a
large fraction on stars on tangential orbits, which in turn requires a
larger enclosed mass to fit the large velocity dispersion.

The fit for the pseudo bulge barred sample gives $(\alpha,
~\epsilon_0) = (7.68\pm0.07, 0.14)$. The offset is $\delta
\alpha=-0.73 \pm 0.09$~dex, which is considerably larger than in the
simulations.  It is unclear whether compression of the bulge by bar
evolution is the main cause for the offset in the case of pseudo
bulges.  The Milky Way, which has a bar and which we have classified
as having a pseudo bulge, is right in the middle of the \Msige\
relation of pseudo bulged barred galaxies.


\section{Discussion \& Conclusions}
\label{sec:conclusions}

\subsection{Offset and scatter in the \Msige\ relation}

We have studied the consequences of angular momentum redistribution
driven by bars on the evolution of the velocity dispersion, \sige, of
the bulge and the implications for the \Msige\ relation.  We showed
that if \Mbh\ does not grow during the formation and evolution of
bars, then the increase in \sige\ results in an offset below the
\Msige\ relation.  The simulations predict an offset $\delta\alpha
\sim -0.2$.

Defining a sample of observed classical bulges from the literature, we
fit the \Msige\ relation of unbarred galaxies.  Then fixing the slope
of the relation, we fit the relation for the classical bulges in
barred galaxies, and find an offset from the unbarred galaxies of
$\delta \alpha = -0.19 \pm 0.20$, consistent with the prediction but
also consistent with no offset.

Contamination of the bulge velocity dispersion by the kinematics of
the disc can lead to changes in \sige\ by as much as $25\%$,
equivalent to an offset in the \Msige\ relation as large as $\delta
\alpha \sim -0.4$ (for $\beta = 4$).  However this contamination
should also be present in the sample of unbarred galaxies relative to
which we measure the offset for the barred galaxies.  Thus
contamination by the disc is very unlikely to cause an offset.

The models imply that the scatter should increase (in quadrature) by
$\sim 0.11$; we measure a scatter for unbarred galaxies of 0.14 and
for barred classical bulges of 0.42.  This is larger than the
predicted scatter, but our prediction is based on a narrow range of
models and does not take into account the possibility that SMBHs can
grow back onto the \Msige\ relation, both of which would increase the
scatter, and considers a narrow range of bulge-to-disc ratios.
Moreover, the observational scatter probably includes a significant
component from modelling uncertainties \citep{Brown2013}.

\subsection{The black hole fundamental plane}

Several studies have suggested that departures from the \Msige\
relation correlate with a third, structural, parameter, such as \re\
or the stellar mass of the bulge \Mbul\ \citep{Marconi2003,
  Francesco2006, Aller2007, Barway2007, Hopkins2007a}.  This has
become known as the black hole fundamental plane (BHFP) and both its
existence and origin have been subject of uncertainty.  This is
because the BHFP, if it exists, is strongly dominated by \sige\
\citep[e.g.][]{Beifiori2012}.  \citet{Hopkins2007b} proposed that the
BHFP may arise from the higher gas mass fraction of merger progenitors
at high redshift.  \citet{Graham2008b} instead argued that barred
galaxies wholly accounted for the BHFP.  We have shown that the
residuals in the \Msige\ relation caused by bar evolution correlate
with structural and kinematic properties of the system.  In the former
case this can account for the weak BHFP measured thus far.  For the
models, we find a strong correlation between $\delta\log\Mbh$ and
${\rm M\B / M\BD}$; observationally however the BHFP is much weaker
and this perhaps reflects the fact that the models have a relatively
narrow range of $B/D$ initially, leading to a strong correlation
between ${\rm M\B / M\BD}$ and $\Delta{\rm M\BD / M\BD_{init}}$.  A
wider range of initial bulge-to-disc ratios is likely to blur the
correlation between ${\rm M\B / M\BD}$ and $\Delta{\rm M\BD /
  M\BD_{init}}$, making for a weaker structural BHFP. In addition, we
are able to fully disentangle bulge and disc in the simulations,
allowing us to compute ${\rm M\B / M\BD}$.  Observationally
disentangling the bulge mass at small radii, where the disc profile
may no longer follow an inward extrapolation of an exponential
profile, may present difficulties.

We have also shown that $\delta\log\Mbh$ strongly correlates with
$\beta_\phi\BD$ and $\beta_z\BD$, which potentially present new
versions of the BHFP where the third parameter is a kinematic one.
This correlation is unlikely to be as sensitive to a wider range of
initial conditions, but this still needs to be tested further.

\subsection{The role of gas}

Using {\it Hubble Space Telescope} STIS spectra to measure upper
limits on \Mbh\ in 105 low-luminosity AGN, \citet{Beifiori2009} found
no offset between the \Msige\ relations of barred and unbarred
galaxies.  Likewise in a study of 76 active galaxies \citet{Xiao2011}
also found no difference between barred and unbarred galaxies.
The main difference between these observations and our results is the
presence of gas.  The simulations presented here are all
collisionless.  As the bar grows, \sige\ increases and SMBHs fall
below the \Msige\ relation.  This offset can be reversed if the SMBH
can grow, which they can best do by accreting gas.  It is now clear
that low to medium luminosity AGN are overwhelmingly resident in disc
galaxies.  Thus secular processes in disc galaxies must play an
important role in the growth of SMBHs \citep{Schawinski2011,
  Cisternas2011, Schawinski2012, Treister2012, Kocevski2012,
  Simmons2012, ArayaSalvo2012, Debattista2013}.  It seems likely that,
after a bar forms, a SMBH will drop below the \Msige\ relation, but,
once gas is driven to the centre, the SMBH can grow again.  If SMBH
growth is governed by AGN feedback, then it would be able to return to
the \Msige\ relation.  This path to returning to the \Msige\ relation
is however not available to galaxies without gas to trickle down to
the SMBH.  The fact that samples with ongoing AGN activity, such as in
\citet{Beifiori2009} and \citet{Xiao2011}, do not show an offset
suggests that bars are efficient at feeding SMBHs.

\subsection{Future observational prospects}

The sample of observed barred galaxies with classical bulges we have
used here includes just five galaxies.  The most immediate way of
extending our results will come from careful classification of the
barred sample of galaxies without bulge classifications.  We explored
what would happen if the sample of unclassified bulge barred galaxies
in Table~\ref{tab:bhmassamples} all hosted classical bulges, which is
very unlikely but gives us an indication of how the offset is likely
to vary.  Fitting the \Msige\ relation with $\beta = 3.82$ gives a
larger offset $\delta\alpha = -0.53 \pm 0.15$ (see
Table~\ref{tab:obsfits} for full fit).  Curiously, other than
NGC~4151, all the rest of these galaxies are offset below the \Msige\
relation, suggesting that a large offset is likely.  Thus the presence
of an offset between unbarred and barred galaxies with classical
bulges may get stronger.

\subsection{Caveats}

Two important caveats need to be borne in mind about our results.
First of all the models considered in this paper have been drawn from
a probability distribution appropriate for properties of the Milky
Way.  At best only one of these models is an accurate representation
of the Milky Way.  It is unlikely that a distribution of models of a
single galaxy is a reasonable representation of the intrinsic variety
of galaxies in general, even at fixed galaxy mass.  For example the
bulge-to-disc ratio in the models takes on a narrow range of values
$0.15 \leq B/D \leq 0.28$, whereas the barred sample in
Table~\ref{tab:bhmassamples} has an order of magnitude larger
variation in $B/D$.  This may bias the values of the offset in the
\Msige\ relation to larger values while decreasing the scatter of the
models.

In addition all the models as constructed are already bar unstable
from the start.  We note in particular that about half the models have
a minimum Toomre-$Q$ between 1.0 and 1.5.  Thus many of the models
need to shed a significant amount of angular momentum in order to form
a bar.  Whether nature forms disc galaxies that are this unstable is
unclear; for instance the high resolution models of
\citet{Roskar2012}, in which the stars all formed out of cooling gas,
rather than put in ab initio as here, tend to evolve at constant $Q$
slightly lower than 2.  Indeed the values in Table~\ref{tab:sims} show
that smaller values of minimum $Q$ produce larger values of $\left<
  \sige/\sigma_{e0}\right>$.  We note that the correlation of $Q$ with
$\left< \sige/\sigma_{e0}\right>$ ($r_s = -0.57$) is stronger than
with \Abar\ ($r_s = 0.19$) or with $\Delta {\rm M/M_{init}}$ ($r_s =
-0.44$).  Thus these models may shed more angular momentum from the
disc centre than in nature, leading to a larger increase in the disc
mass at the centre, a larger increase in \sige\ and thus a larger
offset in the \Msige\ relation.

\subsection{Summary}

We have studied the consequences of bar formation and evolution on the
\Msige\ relation of SMBHs. Our main results can be summarised as
follows:

\begin{itemize}

\item Bars cause an increase in the central mass density of a galaxy,
  altering the kinematics of the bulge and of the disc.  Of particular
  importance for the \Msige\ relation is the increase in \sige.  We
  find a strong correlation between the ratio of final to initial
  dispersion, $\left< \sige/\sigma_{e0} \right>$, and the fractional
  change in mass of the bulge$+$disc within \re\ of the bulge, in good
  agreement with \citet{Debattista2013}.  The simulations show that
  \sige\B\ can increase by as much $\sim 20\%$.  A SMBH in such a
  galaxy would need to grow by a factor of $\sim 2$ to remain on the
  \Msige\ relation.  The average fractional increase of \sige\B\ in
  the simulations is $\left< \sige/\sigma_{e0} \right> = 1.12 \pm
  0.05$.

\item While $\sigma$\BD\ correlates with $\sigma$\B, the two are not
  equal; thus the disc contaminates the measurement of the bulge
  velocity dispersion. In the edge-on view, \sige\BD\ and
  \sigeight\BD\ are up to 25\% larger than \sige\B\ and \sigeight\B.
  $\sigeight\BD/\sigeight\B$ and $\sige\BD/\sige\B$ follow the same
  distribution, but the scatter in \sigeight\BD\ is slightly larger
  than the scatter in \sige\BD. Thus \sige\ is a better quantity for
  studying SMBH scaling relations.

\item We use the \Msige\ relation of \citet{Gueltekin2009b} and the
  models to estimate the offset of barred galaxies in the absence of
  SMBH growth.  We predict an offset $\delta\alpha \sim -0.2$ and an
  increase in scatter by $\eps \sim 0.1$ (in quadrature).

\item We showed that the tangential anisotropy, $\beta_\phi(B+D)$
  correlates very strongly with the change in mass within \re.  Since
  this in turn correlates with the change in \sige, this suggests that
  residuals of galaxies from the \Msige\ relation may also correlate
  very strongly with $\beta_\phi(B+D)$, which is the case for the
  simulations.  This may provide a new version of the black hole
  fundamental plane, where the third parameter is a kinematic one.

\item We use a sample of twelve galaxies to measure the \Msige\
  relation of unbarred disc galaxies with classical bulges.  We find
  $(\alpha,\beta) = (8.41\pm0.05,3.82\pm0.28)$.  Then fixing the slope
  $\beta$, we fit the \Msige\ relation for five barred galaxies with
  classical bulges.  We find $\delta\alpha = -0.19\pm0.20$, comparable
  to the prediction from the models but also consistent with no
  offset.  The same exercise for nine pseudo bulges in barred galaxies
  yields an offset $\delta\alpha = -0.73 \pm 0.09$.  The scatter in
  the \Msige\ relation of the barred classical bulges is larger than
  the one for unbarred classical-bulged galaxies by a amount larger
  than predicted.  This may be because the scatter in the models
  underestimates the real scatter and because SMBHs in barred galaxies
  are able to grow again, returning to the fiducial \Msige\ relation.
  SMBH mass measurements in barred galaxies may also be more uncertain
  than in unbarred galaxies \citep{Brown2013}.

\end{itemize}

\section{Acknowledgements} 

We thank Helen Cammack, Samuel Heald and especially Lindsey Tate, who
worked on parts of the analysis of these simulations during summer
internships. We thank the Nuffield Foundation for supporting Samuel
Heald via a Nuffield Internship during the summer of 2010. Markus
Hartmann thanks Kayan Gueltekin and Alessandra Beifiori for helpful
discussion.  We thank John Dubinski for sharing with us the carefully
constructed simulations used in this paper.  VPD and DRC are supported
by STFC Consolidated grant ST/J001341/1.  MV is supported by U.S.
National Science Foundation grant AST-0908346.  LMW was supported by a
Discovery Grant with the Natural Sciences and Engineering Research
Council of Canada.  The final stages of this work were supported by
the National Science Foundation under Grant No.  PHY-1066293 and the
hospitality of the Aspen Center for Physics.


\bibliographystyle{mn2e1}
\bibliography{smbh}{}


\end{document}